\documentclass[nofootinbib,aps,11pt]{revtex4-1}
\usepackage{graphicx}
\usepackage{hyperref}
\usepackage{cancel}
\usepackage{amssymb}
\usepackage{textcomp}
\usepackage{amsmath}
\usepackage{bm}
\usepackage{times}
\usepackage{epsfig}
\usepackage{color}
\begin{document}
\title{\Large Lepton Number Violation Higgs Decay at Muon Collider}
\bigskip
\author{Fa-Xin Yang$^1$}
\author{Feng-Lan Shao$^1$}
\email{shaofl@mail.sdu.edu.cn}
\author{Zhi-Long Han$^2$}
\email{sps\_hanzl@ujn.edu.cn}
\author{Fei Huang$^2$}
\author{Yi Jin$^{2,3}$}
\author{Honglei Li$^2$}
\email{sps\_lihl@ujn.edu.cn}
\affiliation{
	$^1$School of Physics and Physical Engineering, Qufu Normal University, Qufu, Shandong 273165, China\\
	$^2$School of Physics and Technology, University of Jinan, Jinan, Shandong 250022, China
	\\
	$^3$Guangxi Key Laboratory of Nuclear Physics and Nuclear Technology, Guangxi Normal University, Guilin, Guangxi 541004, China
}
\date{\today}

\begin{abstract}
In this paper, we consider the scalar singlet extension of type-I seesaw, where a scalar singlet $S$ and a heavy neutral lepton $N$ are further introduced. The Majorana mass term of heavy neutral lepton is generated through the Yukawa interaction with the scalar singlet, which then induces the lepton number violation decays of SM Higgs $h$ and heavy Higgs $H$ via mixing of scalars. As a pathway to probe the origin of heavy neutral lepton mass,  we investigate the lepton number violation Higgs decay signature  at the TeV-scale muon collider. The dominant production channel of Higgs bosons at the TeV-scale muon collider is via vector boson fusion. So we perform a detailed analysis of the signal process $\mu^+\mu^-\to \nu_\mu \bar{\nu}_\mu  h/H \to \nu_\mu \bar{\nu}_\mu NN $ followed by $ N \to \mu^\pm jj$, where
the two jets from $W$ boson decay are treated as one fat-jet $J$. With an integrated luminosity of $1(10)~\text{ab}^{-1}$, the 3 (10) TeV muon collider could discover the lepton number violation SM Higgs decay $h\to \mu^\pm\mu^\pm JJ$ signature for the Higgs mixing parameter $\sin\alpha>0.05(0.009)$. Meanwhile, a large parameter space can be detected by the lepton number violation heavy Higgs decay $H\to \mu^\pm\mu^\pm JJ$ signature for $m_H\lesssim1 (3)$~TeV and $\sin\alpha\gtrsim0.03(0.005)$ at the 3 (10) TeV muon collider. Therefore, the lepton number violation SM and heavy Higgs decay signatures are both promising at the TeV scale muon collider.
	
\end{abstract}

\maketitle

%%%%%%%%%%%%%%%%%%%%%%%
\section{Introduction}
%%%%%%%%%%%%%%%%%%%%%%%

The discovery of the 125 GeV Higgs boson at the large hadron collider (LHC) is a great achievement of the standard model (SM) \cite{ATLAS:2012yve,CMS:2012qbp}, which also provides a unique pathway to probe new physics beyond SM. Meanwhile, the neutrino oscillation experiments indicate tiny but non-zero neutrino masses \cite{Super-Kamiokande:1998kpq,SNO:2002tuh,DayaBay:2012fng}. If neutrinos are Dirac particles, their masses could originate from the Yukawa interaction with the SM Higgs doublet $\Phi$ as $y_D\bar{L}_L \tilde{\Phi}\nu_R$. To explain the sub-eV neutrino masses, $y_D\lesssim10^{-12}$ is required, which is theoretically unnatural small compared to other SM Yukawa couplings.

One natural pathway to explain the tiny neutrino mass is the seesaw mechanism \cite{Minkowski:1977sc,Mohapatra:1979ia,Schechter:1980gr,Schechter:1981cv}, which introduces heavy neutral lepton $N$. Due to the Majorana nature of $N$, one distinct signal is the lepton number violation signature at the collider \cite{delAguila:2005ssc,Banerjee:2015gca,Drewes:2019byd,Mekala:2022cmm,Antusch:2022ceb,Mikulenko:2023ezx,Mekala:2023kzo}, such as $pp \to W^{\pm(*)}\to \ell^\pm N \to \ell^\pm \ell^\pm jj$ \cite{Han:2006ip,Deppisch:2015qwa,Cai:2017mow}. In the well-studied phenomenological type-I seesaw model, the lepton number violation signature is determined by the mixing parameter $V_{\ell N}$ between the light and heavy neutrinos \cite{Antusch:2016ejd,Abdullahi:2022jlv},  which is mostly sensitive to the mass region with $m_N\lesssim100$ GeV. Above the electroweak scale, $V_{\ell N}$ is less constrained as $|V_{\ell N}|^2\gtrsim10^{-3}$ being disfavored via indirect constraints \cite{Antusch:2015mia}. Currently, the LHC searches could exclude $|V_{\ell N}|^2\gtrsim2\times10^{-6}$ at $m_N=20$ GeV for prompt \cite{ATLAS:2019kpx,CMS:2024xdq} and $|V_{\ell N}|^2\gtrsim 3\times10^{-7}$ at $m_N=12$ GeV for displaced signature \cite{ATLAS:2022atq,CMS:2022fut,CMS:2023jqi}. In the future, CEPC and FCC-ee could probe $|V_{\ell N}|^2\gtrsim10^{-10}$ with $m_N\sim$50 GeV \cite{Abdullahi:2022jlv}, and the muon collider could test $|V_{\ell N}|\gtrsim10^{-6}$ for TeV scale $m_N$ \cite{Mekala:2023diu,Kwok:2023dck,Li:2023tbx}. However, the natural theoretical predicted squared mixing parameter is $|V_{\ell N}|^2\sim m_\nu/m_N<10^{-10}$ for $m_N>1$ GeV, which is even beyond the reach of future experiments \cite{Abdullahi:2022jlv}.

The Majorana mass term $m_N \bar{N}^c N$ is allowed by the SM gauge symmetry, so $m_N$ is usually assumed to be large enough to suppress the neutrino mass in the canonical seesaw. If we assign the lepton number $L=1$ to $N$, the Majorana mass term clearly breaks the lepton number symmetry by two units. Since $m_N\to 0$ will restore the lepton number symmetry, the bare Majorana mass term should be theoretically preferred to be small \cite{tHooft:1979rat}, just such as in the inverse seesaw mechanism \cite{Mohapatra:1986bd}. On the other hand, the origin of the Majorana mass term is not clarified in the minimal seesaw either. This term can be realized by extending the SM gauge symmetry, such as the left-right symmetrical model \cite{Pati:1974yy,Mohapatra:1974gc,Senjanovic:1975rk}, the gauged $U(1)_{B-L}$ model \cite{Davidson:1978pm,Marshak:1979fm,Mohapatra:1980qe,Liu:2021akf}, the gauged $U(1)_{L_\mu-L_\tau}$ model \cite{Branco:1988ex,He:1990pn,Asai:2017ryy,He:2024dwh,Bi:2024pkk}, and the general gauged $U(1)_X$ model \cite{Cox:2017eme,Das:2017flq,Das:2017deo,Das:2022rbl,Arun:2022ecj}. In such gauge symmetry extended models, heavy neutral lepton $N$ has additional interactions with the new gauge boson and new scalars. Searches of heavy neutral lepton via decays of new gauge bosons have been performed by LHC \cite{ATLAS:2023cjo,CMS:2023ooo}. For instance, the search for pair production of heavy neutral lepton via the process $pp\to Z'\to NN$ with $N\to \ell^\pm jj$ by the CMS collaboration has excluded the parameter space with $m_{Z'}\lesssim4.4$ TeV and $m_N\lesssim1.3$ TeV \cite{CMS:2023ooo}.  In the future, the HL-LHC could probe the parameter space with $m_{Z'}\lesssim6$ TeV and $m_N\lesssim3$ TeV \cite{Arun:2022ecj}.

In this paper, we focus on the Higgs sector with an additional scalar singlet $S$ \cite{Shoemaker:2010fg,Gao:2019tio}, which usually appears in the $U(1)$ symmetric models. By assigning lepton number $L=-2$ to the new scalar $S$, the Majorana mass term then originates from the new Yukawa interaction $S \bar{N}^c N$, which is invariant under the $U(1)$ symmetry. After the spontaneous symmetry breaking, both the scalar singlet and the neutral component of SM Higgs doublet $\Phi$ obtain non-zero vacuum expectation values. The scalar interaction $(\Phi^\dag \Phi) (S^\dag S)$ induces the mixing between scalars. In the mass eigenstate, we denote $h$ to be the SM 125~GeV Higgs boson and $H$ to be the additional heavy Higgs boson. Including the mixing between scalars, both the SM $h$ and heavy Higgs boson $H$ couple to the heavy neutral lepton $N$, which could induce the lepton number violation Higgs decay as $h/H\to NN $ with $ N \to \ell^\pm jj$ at colliders when kinematically allowed, i.e., $m_N<m_h/2$ or $m_N<m_H/2$ \cite{Maiezza:2015lza}.

 The lepton number violation Higgs decay has been investigated at hadron colliders \cite{Basso:2010yz,Nemevsek:2016enw,Accomando:2017qcs,Fuks:2025jrn,Liu:2025ldf}, which is sensitive to the parameter space with $m_H\lesssim150$ GeV \cite{Nemevsek:2016enw}. Meanwhile, the future $e^+e^-$ colliders could probe the mixing parameter between scalars $\sin\alpha$ to be lower than $10^{-2}$\cite{Gao:2021one,Barducci:2020icf}.  In this paper, we study this signature at the recently proposed muon collider \cite{Delahaye:2019omf,Accettura:2023ked,InternationalMuonCollider:2024jyv,InternationalMuonCollider:2025sys}. Compared to the hadron collider, the muon collider has a cleaner environment. Meanwhile, the muon collider can achieve much higher center-of-mass energy than the $e^+e^-$ collider \cite{MuCoL:2024oxj}, which also enables the reach of the heavy Higgs boson. Different from searches for heavy neutral lepton in canonical seesaw \cite{Mekala:2023diu,Kwok:2023dck,Li:2023tbx,Li:2023lkl,Cao:2024rzb}, this signature is not suppressed by the mixing parameter $V_{\ell N}$. Meanwhile, instead of the  Higgs-strahlung process at the Higgs factory \cite{Gao:2021one}, the dominant production channel of Higgs bosons at the TeV scale muon collider is via vector boson fusion \cite{Han:2020pif,Andreetto:2024rra}, which provides a new independent pathway to probe the intrinsic of Higgs bosons. So the full signal process investigated in this paper is $\mu^+\mu^-\to \nu_\mu \bar{\nu}_\mu  h/H \to \nu_\mu \bar{\nu}_\mu NN $ followed by $ N \to \ell^\pm jj$.

 Besides the neutral Higgs bosons $h/H$,  there could be other origin of the lepton number violation signature at the muon collider, such as the new gauge boson $Z'$ \cite{He:2024dwh} and the charged Higgs boson $H^\pm$ \cite{Batra:2023ssq}. The common same sign dilepton signature $\ell^\pm\ell^\pm 4j$ could be induced as $\mu^+\mu^-\to \nu_\mu \bar{\nu}_\mu  h/H \to \nu_\mu \bar{\nu}_\mu NN \to \ell^\pm\ell^\pm 4j +  \nu_\mu \bar{\nu}_\mu $ via the neutral Higgs bosons $h/H$, $\mu^+\mu^-\to Z^{\prime *}\to NN \to \ell^\pm \ell^\pm 4j$ via the gauge boson $Z'$, or $\mu^+\mu^-\to NN\to \ell^\pm \ell^\pm 4j $ via the $t$-channel exchange of charged Higgs boson $H^\pm$.  These origins can be distinguished via additional signatures. For instance, the new gauge boson could mediate the new signature $\mu^+\mu^-\to Z' \gamma \to NN\gamma\to \ell^\pm \ell^\pm 4j+ \gamma$ \cite{He:2024dwh}.  Meanwhile the charged Higgs boson can be pair produced at muon collider, which leads to the new signature as $\mu^+\mu^-\to H^+H^-\to \ell^+N+\ell^-N\to \ell^+\ell^- \ell^\pm\ell^\pm 4j$ \cite{Batra:2023ssq}.

The rest of the paper is organized as follows. In Section \ref{SEC:TM}, we briefly review the scalar singlet extension of the Type-I seesaw model. Constraints on the heavy Higgs boson are also discussed. The decay properties of scalars $h/H$ and heavy neutral lepton $N$ are considered in Section \ref{Sec:DP}. The lepton number violation signatures from SM Higgs and heavy Higgs at the TeV-scale muon collider are analyzed in Section \ref{Sec:SM} and Section \ref{Sec:HH}, respectively. The conclusion is in Section \ref{SEC:CL}.

\section{The Model}\label{SEC:TM}

The studied model in this paper is the type-I seesaw with an additional scalar singlet $S$. This singlet $S$ usually appears in $U(1)$ extension of seesaw models. To explain the smallness of neutrino mass,  the seesaw mechanism works with heavy neutral lepton $N$ \cite{Minkowski:1977sc,Mohapatra:1979ia,Schechter:1980gr,Schechter:1981cv}, which carries lepton number $+1$. The scalar singlet $S$ with lepton number $-2$ not only breaks the $U(1)$ symmetry spontaneously, but also generates the Majorana mass of heavy neutral lepton via the Yukawa interaction $S\bar{N}^c N$. Hence, the mass of heavy neutral lepton $m_N$ is strongly associated with the scale of $U(1)$ symmetry, which we attribute to the TeV scale for collider searches. The local $U(1)$ scenario is considered in this paper to avoid the massless Goldstone boson \cite{Chikashige:1980qk,Pilaftsis:1993af}. For different $U(1)$ symmetry, it can be distinguished from the interaction of new gauge boson due to different charge assignment of fermions \cite{KA:2023dyz}, while the scalar sectors are similar in the $U(1)$ models.

Under the additional $U(1)$ symmetry, the scalar potential of SM Higgs doublet $\Phi$ and scalar singlet $S$ is given by 
\begin{align}
	V(\Phi,S)=m_1^2 \Phi^\dagger \Phi+m_2^2 S^\dag S+\lambda_1(\Phi^\dagger \Phi)^2+\lambda_2 (S^\dag S)^2+\lambda_3(\Phi^\dagger \Phi)(S^\dag S),
\end{align}
where $m_1^2<0,m_2^2<0$ is required to break the symmetry spontaneously. The stability
of the potential is satisfied with the condition
\begin{equation}
	\lambda_1>0,~\lambda_2>0, ~4\lambda_1 \lambda_2 - \lambda_3^2>0.
\end{equation}

After the spontaneous symmetry breaking, the neutral component of scalars obtain vacuum expectation values as $\langle\Phi\rangle= v_0/\sqrt{2}$ and $\langle S \rangle= v_s/\sqrt{2}$, where $v_0=246$ GeV. In the following studies, we fix $v_s=1$~TeV for illustration. The $\lambda_3 (\Phi^\dagger \Phi)(S^\dag S)$ term then induces mixing between the scalars. The mass matrix for the two CP-even Higgs bosons $\phi^0$ and $s^0$ is given by
\begin{align}
	M^2(\phi^0,s^0)=\begin{pmatrix}
		2\lambda_1v_0^2 & \lambda_3v_0v_s \\ 
		\lambda_3v_0v_s & 2\lambda_2v_s^2 
	\end{pmatrix}.
\end{align}
 The mass eigenstate can be acquired by an orthogonal transformation
 \begin{align}
 	\begin{pmatrix}
 		h\\
 		H
 	\end{pmatrix}
 	=\begin{pmatrix}
 		\cos\alpha & -\sin\alpha \\ 
 		\sin\alpha & \cos\alpha 
 	\end{pmatrix}
 	\begin{pmatrix}
 		\phi^0\\
 		s^0
 	\end{pmatrix},
 \end{align}
 where the mixing angle $\alpha$ is derived as
 \begin{align}
 	\tan(2\alpha)=\frac{\lambda_3v_0v_s}{\lambda_2v_s^2-\lambda_1v_0^2}.
 \end{align}
 
 Then, the masses for the two physical Higgs states $h,H$ are
 \begin{align}
 	m_{h}^2=\lambda_1v_0^2+\lambda_2v_s^2-\sqrt{(\lambda_1v_0^2-\lambda_2v_s^2)^2+(\lambda_3v_0v_s)^2},\\
 	m_{H}^2=\lambda_1v_0^2+\lambda_2v_s^2+\sqrt{(\lambda_1v_0^2-\lambda_2v_s^2)^2+(\lambda_3v_0v_s)^2},
 \end{align}
where $h$ is the 125 GeV SM Higgs, and $H$ is the additional heavy Higgs. In this paper, we consider the scenario with $m_H>m_h$.

The mixing between scalars modifies the coupling of SM Higgs by a factor of $\cos\alpha$, which then  alters the Higgs signal strengths.  In this paper, we consider the constraints from current LHC searches. Using the precision Higgs measurements data \cite{ATLAS:2022vkf,CMS:2022dwd}, the global fit result of all channels requires $|\sin\alpha|<0.29$ \cite{Lane:2024vur,Lewis:2024yvj}. More stringent constraints come from the direct searches for the heavy Higgs in channels of  $H\to ZZ$ \cite{CMS:2018amk,ATLAS:2020tlo}, $H\to WW$ \cite{CMS:2019bnu,ATLAS:2020fry}, and $H\to hh$ \cite{CMS:2021yci,CMS:2022kdx,ATLAS:2021ifb,ATLAS:2022xzm}. The combined analysis of all channels shows that $|\sin\alpha|\lesssim0.2$ is allowed by direct searches \cite{Feuerstake:2024uxs}, where the explicit constraint on $\sin\alpha$ depends on the mass of heavy Higgs.
Meanwhile, the NLO calculations of $W$ boson mass set the most strict constraints for $m_H$ around the TeV scale, which could  exclude $|\sin\alpha|\gtrsim0.16$ \cite{Lopez-Val:2014jva,Robens:2016xkb,Feuerstake:2024uxs}. 

The relevant Yukawa interactions in the scalar singlet extension of the type-I seesaw are
\begin{equation}
	-\mathcal{L}_Y = y_D \bar{L}_L \tilde{\Phi} N + \frac{y_N}{2}  S \bar{N}^c N + \text{h.c.},
\end{equation}
where $L_L$ denotes the left-handed lepton doublet. After the symmetry breaking, the Dirac and Majorana mass terms are generated as $m_D=y_D v_0 /\sqrt{2}$ and $m_N= y_N v_s/\sqrt{2}$. Through the seesaw mechanism,  light neutrino mass is derived as $m_\nu = m_D m_N^{-1} m_D^T$. The mixing parameter between the light and heavy neutrino is
\begin{equation}
	V_{\ell N} = \frac{m_D}{m_N} \sim \sqrt{\frac{m_\nu}{m_N}}.
\end{equation}
Without fine-tuning, the typical value of the mixing parameter is $|V_{\ell N}|^2\sim m_\nu/m_N<10^{-12}$ for heavy neutral lepton above the electroweak scale. In this paper, we fix the mixing parameter as $|V_{\ell N}|^2=m_\nu/m_N=10^{-12}\times (50~\text{GeV})/m_N$ with $m_\nu=0.05$ eV in the following studies. For instance, $m_N=50 (500)$ GeV corresponds to $|V_{\ell N}|^2=10^{-12}(10^{-13})$. Such  small mixing parameters are far beyond the reach of current experimental limits \cite{ATLAS:2019kpx,CMS:2024xdq}. 
 
Including the mixing between light and heavy neutrinos, the interactions of heavy neutral leptons are
\begin{eqnarray}
	\mathcal{L}&\supset& -\frac{g}{\sqrt{2}} W_\mu \bar{N} V_{\ell N}^{*} \gamma^\mu P_L \ell  
	-\frac{g}{2\cos\theta_W} Z_\mu \bar{N} V_{\ell N}^{*}  \gamma^\mu P_L \nu_\ell   - \frac{g m_N}{2 m_W} \cos\alpha~ h \bar{N} V_{\ell N}^{*} P_L \nu_\ell \nonumber \\  
	&& +  \frac{m_N}{2v_s} \sin\alpha~ h \bar{N}^cN-\frac{m_N}{2v_s} \cos\alpha ~H \bar{N}^cN + \text{h.c.},
\end{eqnarray}
where $\theta_W$ is the weak mixing angle.

\section{Decay Properties}\label{Sec:DP}

In principle, the decay mode $h\to N\nu_\ell$ is allowed if $m_N<m_h$. However, this decay mode is suppressed by the tiny natural mixing parameter $V_{\ell N}$, which can be safely neglected in the following discussion. Once kinematically permitted, the new decay mode is $h\to NN$. The corresponding partial decay width is
\begin{align}\label{Eqn:hNN}
	\Gamma(h\rightarrow NN)=\frac{m_N^2 m_h}{4\pi v_s^2} \sin^2\alpha \left(1-\frac{4m_N^2}{m_h^2}\right)^{3/2},
\end{align}
which then leads to the branch ratio 
\begin{align}
	{\rm{BR}}(h\rightarrow NN)=\frac{\Gamma(h\rightarrow NN)}{\Gamma(h) \cos^2\alpha+\Gamma(h\rightarrow NN)}.
\end{align}
Here, $\Gamma(h)=4.07$ MeV is the total decay width of SM Higgs \cite{LHCHiggsCrossSectionWorkingGroup:2013rie}.

\begin{figure}
	\begin{center}
		\includegraphics[width=0.45\linewidth]{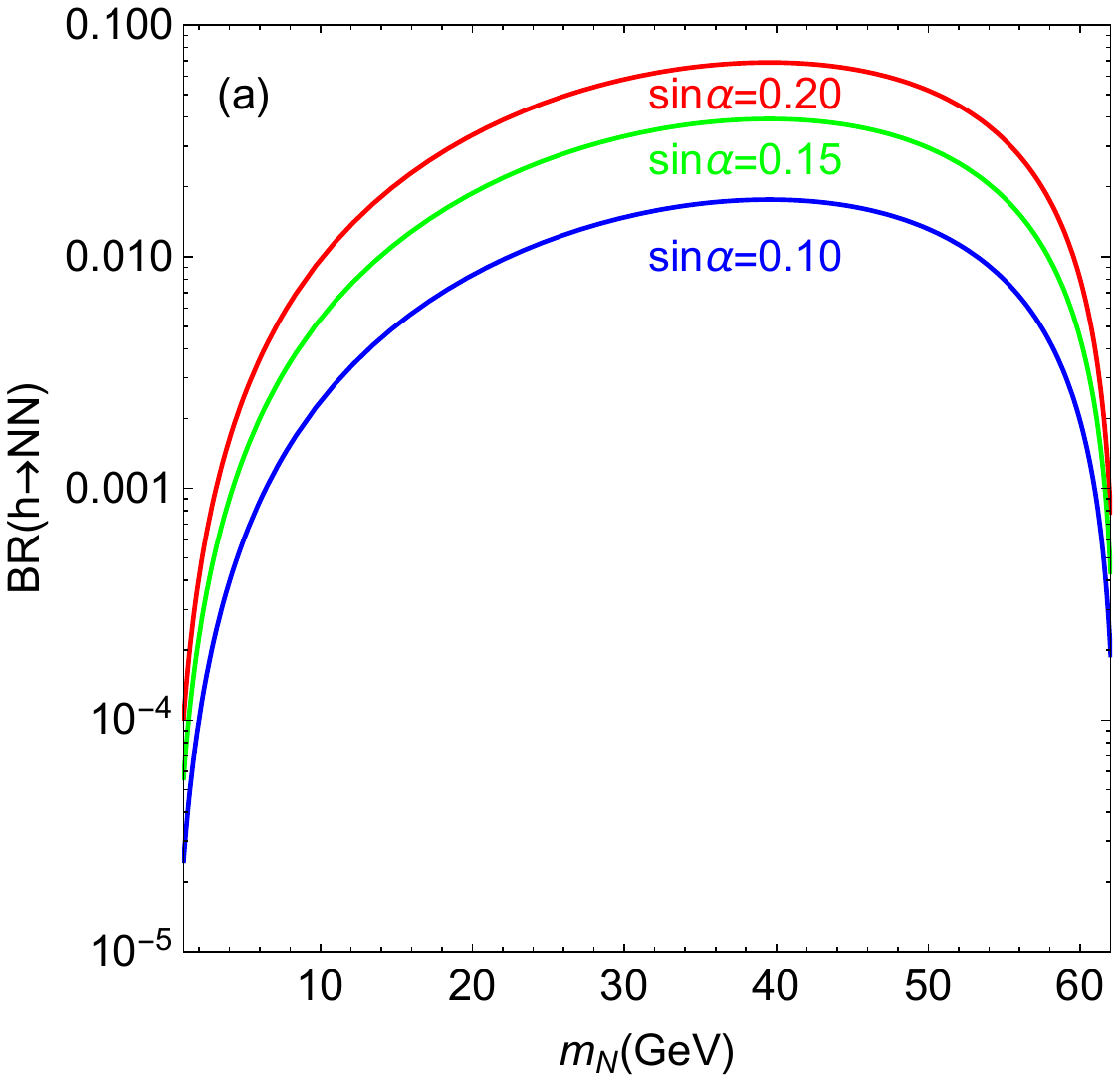}
		\includegraphics[width=0.45\linewidth]{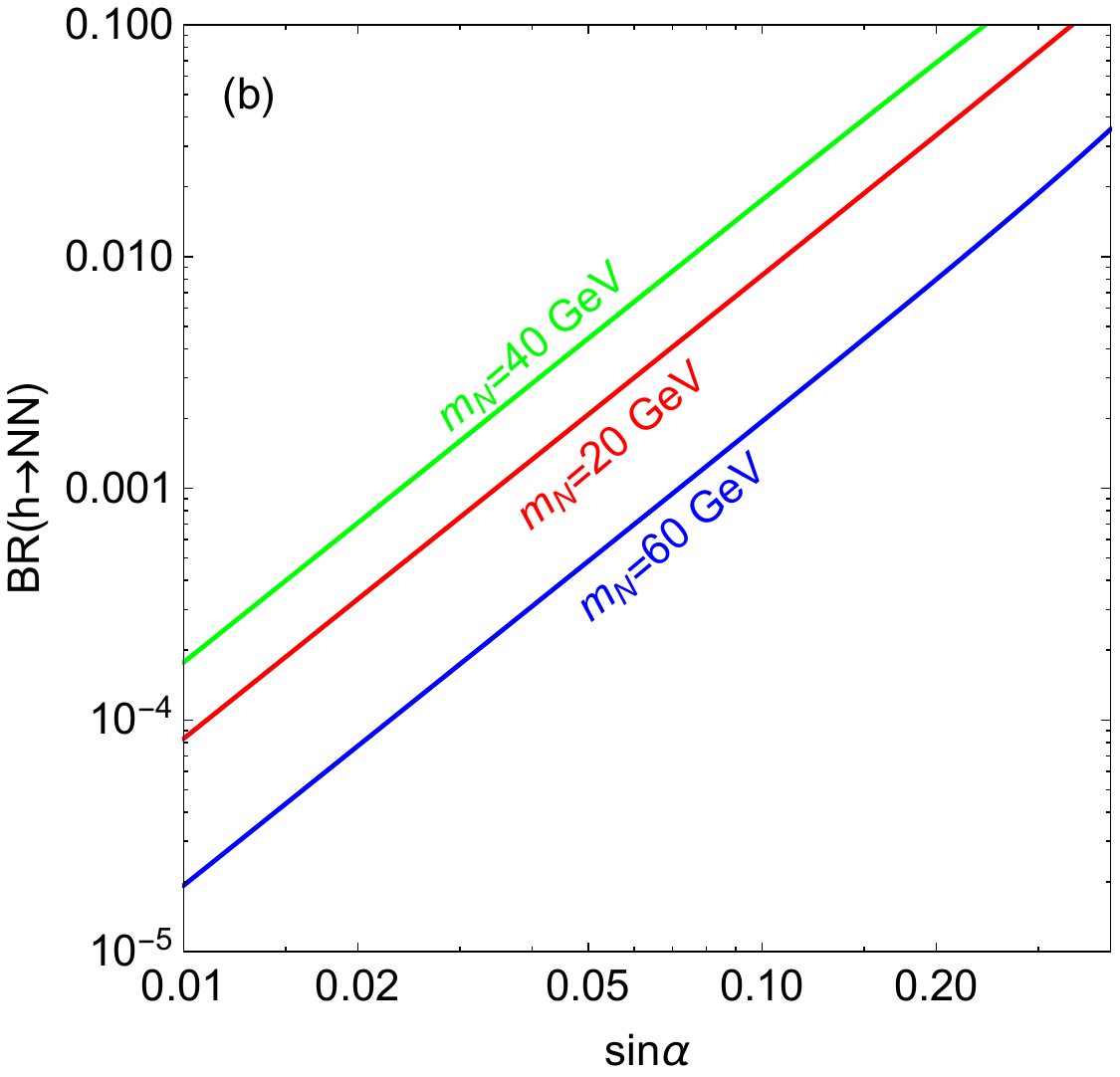}
	\end{center}
	\caption{Branching ratio of $h\to NN$ as a function of $m_N$ in panel (a) and $\sin\alpha$ in panel (b).}
	\label{fig1}
\end{figure}

Theoretical predictions of BR$(h\to NN)$ are shown in Figure \ref{fig1}. With a typical value of $\sin\alpha\sim\mathcal{O}(0.1)$ under current experimental limits, the predicted branching ratio of $h\to NN$ is at the order of 0.01. As we fixed $v_s=1000$ GeV in this paper, decreasing $m_N$ will lead to smaller Yukawa coupling $y_N$, hence a smaller branching ratio. Meanwhile, the heavier $N$ has a smaller phase space of $h\to NN$,  which results in the maximum branching ratio around $m_N=40$ GeV.  It is clear in Equation~\eqref{Eqn:hNN} that the partial decay width  $\Gamma(h\to NN)$ is proportional to $\sin^2\alpha$. When $\sin\alpha\gtrsim0.05$, the predicted BR$(h\to NN)$ is larger than 0.001 for $m_N\sim10-50$ GeV.

The heavy Higgs $H$ couples intrinsically to heavy neutral lepton $N$. Through mixing with SM Higgs $h$, the heavy Higgs $H$ also couples to SM fermions and gauge bosons with the corresponding couplings proportional to $\sin\alpha$. If kinematically allowed, the heavy Higgs also decays into the SM Higgs pair. The partial decay widths of $H$ are calculated as
\begin{eqnarray}\label{Eqn:11}
\Gamma(H\rightarrow NN)&=& \frac{m_N^2 m_H}{4\pi v_s^2} \cos^2\alpha \left(1-\frac{4m_N^2}{m_H^2}\right)^{3/2},\\
\Gamma(H\rightarrow W W )&=&\frac{m_H^3 \sin^2\alpha}{16\pi v_0^2} \left(1-\frac{4m_W^2}{m_H^2}+ \frac{12m_W^4}{m_H^4}\right)\left(1-\frac{4m_W^2}{m_H^2}\right)^{1/2},\\
\Gamma(H\rightarrow ZZ)&=&\frac{m_H^3 \sin^2\alpha}{32\pi v_0^2} \left(1-\frac{4m_Z^2}{m_H^2}+ \frac{12m_Z^4}{m_H^4}\right)\left(1-\frac{4m_Z^2}{m_H^2}\right)^{1/2},\\
\Gamma(H\rightarrow t\bar{t})&=&\frac{3 m_t^2 m_H}{8\pi v_0^2}\sin^2\alpha \left(1-\frac{4 m_t^2}{m_H^2}\right)^{3/2},\\
\Gamma(H\rightarrow hh)&=&\frac{g_{Hhh}^2}{8\pi m_H}\left(1-\frac{4m_h^2}{m_H^2}\right)^{1/2},
\end{eqnarray}
where the cubic coupling $g_{Hhh}$ is determined by
\begin{equation}
	g_{Hhh} = -\frac{\sin 2\alpha}{2 v_0 v_s}(v_0 \sin\alpha + v_s \cos\alpha) \left(m_h^2+\frac{m_H^2}{2}\right).
\end{equation}
In this paper, we consider the heavy Higgs mass larger than 200 GeV, thus the two-body decay modes into SM gauge boson $H\to WW,ZZ$ are always allowed. Theoretically, $m_H<2m_W$ is also possible \cite{Fuks:2025jrn}.
 
\begin{figure}
	\begin{center}
		\includegraphics[width=0.46\linewidth]{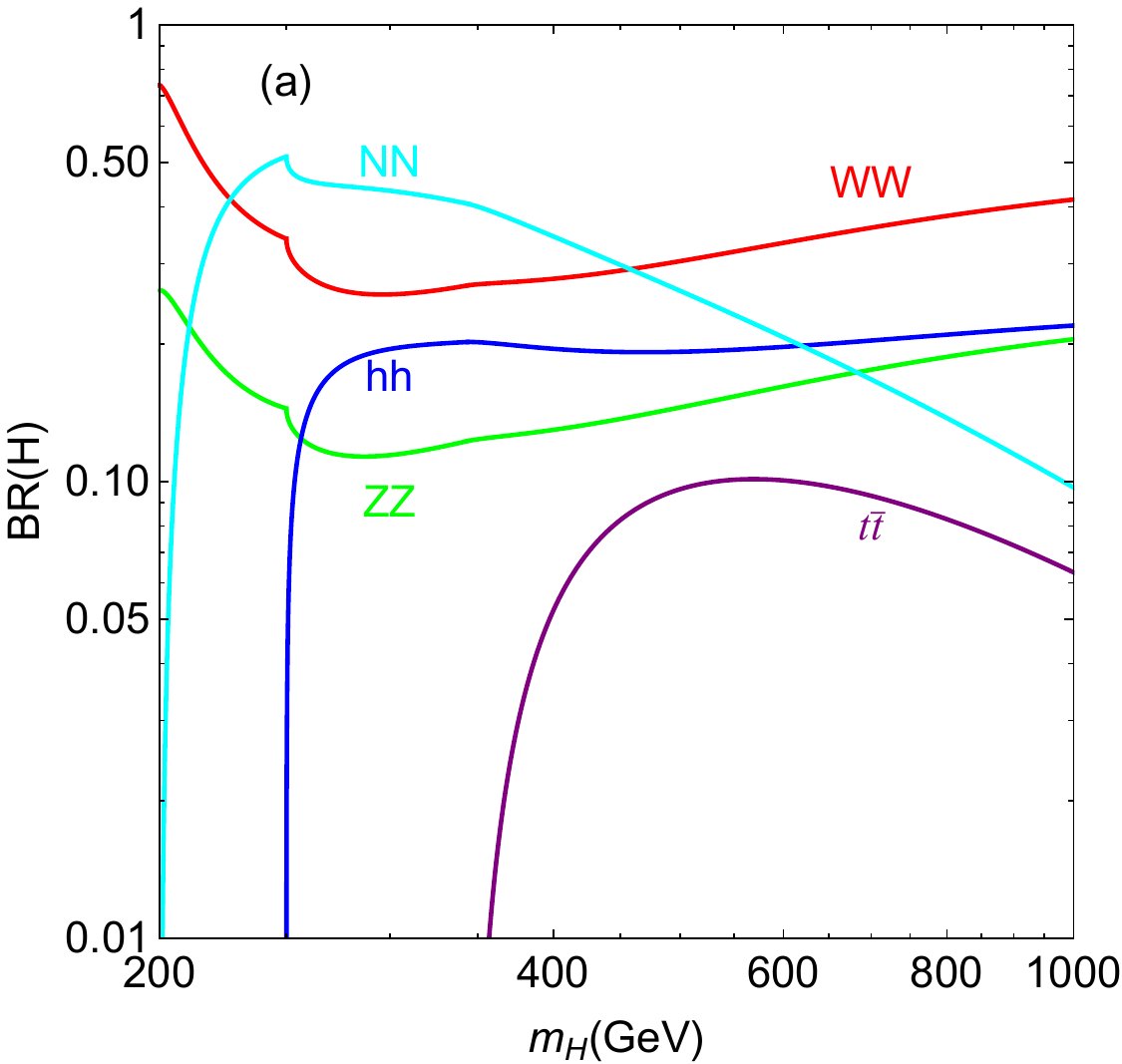}
		\includegraphics[width=0.44\linewidth]{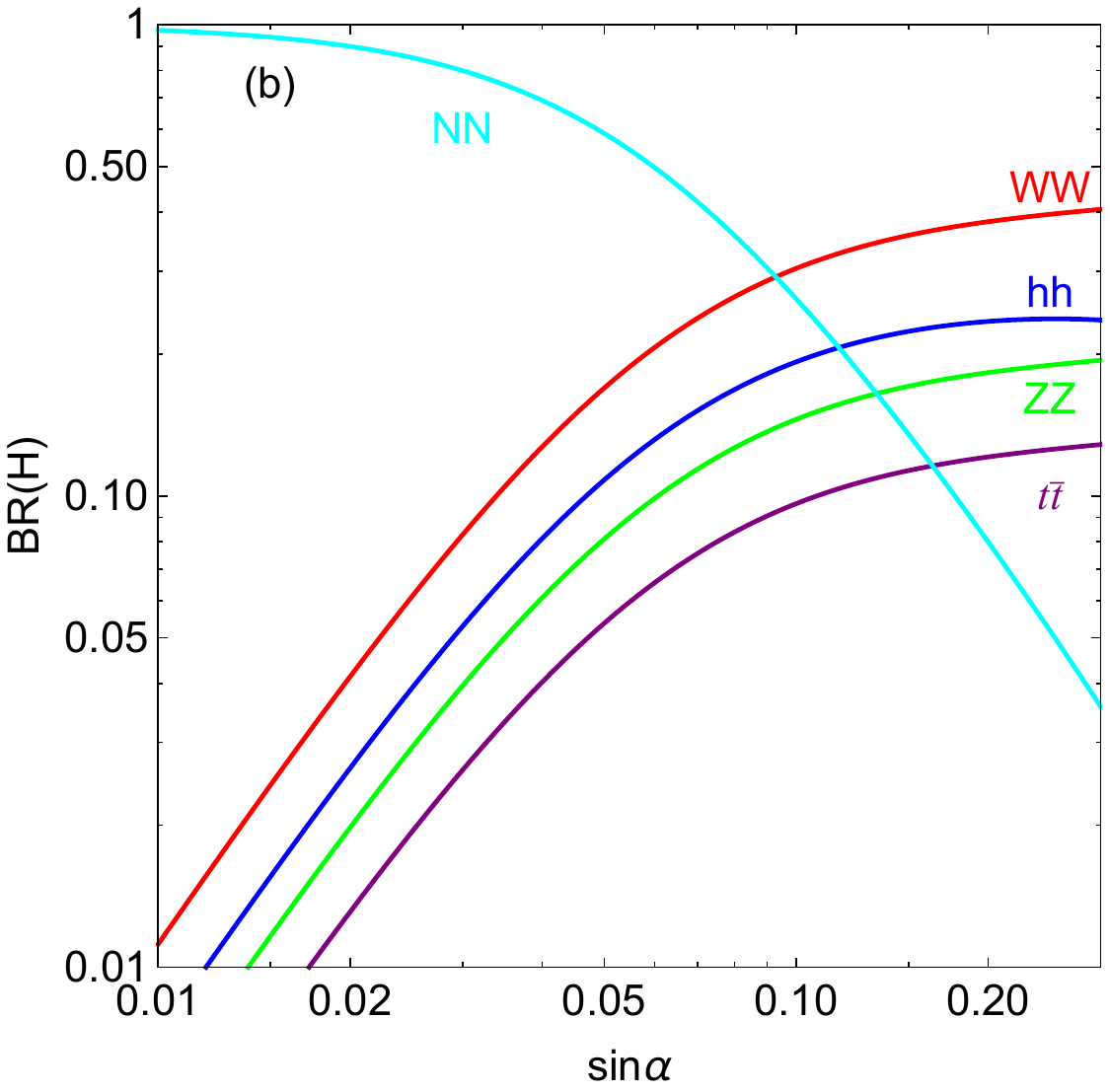}
	\end{center}
	\caption{Branching ratio of heavy Higgs $H$ as a function of $m_H$ in panel (a) and $\sin\alpha$ in panel (b). We have set $m_N=100~\rm{GeV}$ and $\sin\alpha=0.1$ in panel (a). Meanwhile, $m_H$ is fixed to 500 GeV in panel (b).}
	\label{fig2}
\end{figure}

In Figure \ref{fig2}, we show the branching ratio of heavy Higgs. With $\sin\alpha=0.1$, the dominant decay mode is $H\to NN$ in the mass range of [230, 450] GeV, which indicates that the lepton number violation signature could be the best channel for electroweak scale heavy Higgs. We note that the decay widths of heavy Higgs into fermions are proportional to $m_H$, while the decay widths of heavy Higgs into bosons are proportional to $m_H^3$. Therefore, the wanted branching ratio of $H\to NN$ decreases as $m_H$ increases. For heavy Higgs above 600 GeV, we approximately have 
\begin{equation}
	\frac{1}{2}\text{BR}(H\to WW)\approx \text{BR}(H\to hh)\approx\text{BR}(H\to ZZ),
\end{equation}
which reveals the Goldstone nature of gauge bosons.

On the other hand, the couplings of heavy Higgs to SM particles are suppressed by the Higgs mixing angle. Therefore, a smaller mixing angle will lead to a larger branching ratio of $H\to NN$, which is depicted in panel (b) of Figure \ref{fig2}. Typically for $\sin\alpha\lesssim0.1$, the $H\to NN$ channel becomes the dominant decay mode. 

It should be noted that current searches of heavy Higgs are based on the SM final states as $H\to ZZ$, $H\to WW$ and $H\to hh$. The non-observation of such heavy Higgs decay has already set stringent limits on the Higgs mixing angle, i.e., $|\sin\alpha|\lesssim0.2$. Within a certain mass region or small Higgs mixing angle, the dominant decay mode of heavy Higgs becomes $H\to NN$. If there is still no positive signature of heavy Higgs into the SM final states at colliders in the future, the $H\to NN$ channel should be regarded as the golden mode. We have to mention that the production of heavy Higgs $H$ at colliders is induced by the mixing with SM Higgs, thus the cross section is proportional to $\sin^2\alpha$. Although a tiny $\sin\alpha$ leads to $H\to NN$ becoming the dominant channel, it also suppresses the production cross section of heavy Higgs. In Section \ref{Sec:HH}, we will figure out the sensitive region at the TeV scale muon collider.

\allowdisplaybreaks

\begin{figure}
	\begin{center}
		\includegraphics[width=0.45\linewidth]{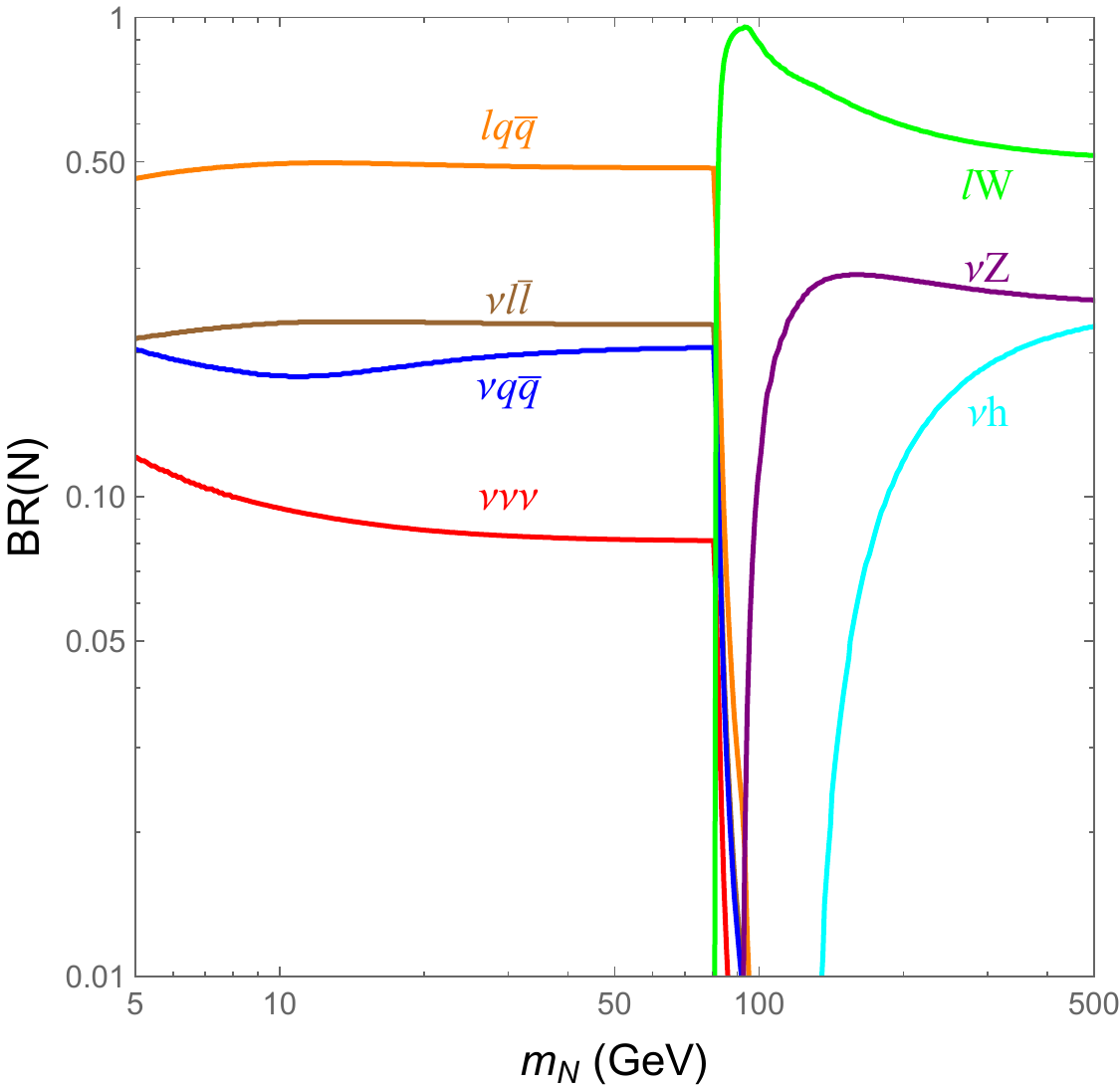}
	\end{center}
	\caption{Branching ratio of heavy neutral lepton $N$. We have fixed $\sin\alpha=0.1$.}
	\label{fig3}
\end{figure}

Decays of heavy neutral lepton $N$ are determined by the mixing parameter with light neutrinos. For sufficient heavy $N$, the two body decay modes $N\to \ell^\pm W^\mp, \nu_\ell Z, \nu_\ell h$ are the dominant channels. The corresponding partial decay widths are 
\begin{eqnarray}
	\Gamma(N\to \ell^\pm W^\mp) &=& \frac{g^2}{64\pi} |V_{\ell N}|^2 \frac{m_N^3}{m_W^2}\left(1-\frac{m_W^2}{m_N^2}\right)^2\left(1+2\frac{m_W^2}{m_N^2}\right), \\
	\Gamma(N\to \nu_\ell Z) &=& \frac{g^2}{128\pi} |V_{\ell N}|^2 \frac{m_N^3}{m_W^2}\left(1-\frac{m_Z^2}{m_N^2}\right)^2\left(1+2\frac{m_Z^2}{m_N^2}\right),\\
	\Gamma(N\to \nu_\ell h) & =& \frac{g^2}{128\pi} |V_{\ell N}|^2 \frac{m_N^3}{m_W^2}\left(1-\frac{m_h^2}{m_N^2}\right)^2 \cos^2\alpha.
\end{eqnarray}

When the heavy neutral lepton is lighter than the $W$ boson, the three body decays  mediated by off-shell $W$ and $Z$ bosons are the dominant decay modes. Analytical expressions for the three body partial decay widths can be found in Ref. \cite{Atre:2009rg,Liao:2017jiz}. In this paper, we use {\bf Madgraph5\_aMC@NLO} \cite{Alwall:2014hca} to evaluate the numerical results of branching ratios, which is shown in Figure \ref{fig3}. The three body decay mode $N\to \ell^\pm q \bar{q}'$ has a branching ratio of about 0.5 when $m_N<m_W$. Meanwhile, the two body decay mode $N\to \ell^\pm W^\mp$ is larger than 0.5 when $m_N>m_W$. Hence, the fully visible  signature $N\to \ell^\pm jj$ is always a dominant channel for $m_N>5$ GeV.

Depending on the explicit decay mode of the heavy neutral lepton, there could be various signal channels from the cascade decays of the $NN$ pair. Focusing on the final states with charged leptons, the explicit signatures could be
\begin{eqnarray}
	NN&\to& \ell^\pm jj + \ell^\pm jj \to 2\ell^\pm +4j, \\
	&\to & \ell^\pm jj + \ell^\mp jj \to \ell^+ \ell^- + 4j, \\
	&\to & \ell^\pm jj + \nu_\ell \ell^+ \ell^- \to 3\ell + 2j + \cancel{E}_T, \\
	&\to & \nu_\ell \ell^+ \ell^- + \nu_\ell \ell^+ \ell^- \to 4\ell +\cancel{E}_T.
\end{eqnarray}
A full study of these signatures is beyond the scope of this paper. As the smoking gun of Majorana neutrino, we investigate the lepton number violation signature $2\ell^\pm +4j$ in this work, which also has relatively small SM backgrounds compared to the lepton number conserving signature.

\section{Signature from SM Higgs}\label{Sec:SM}

We first consider the pair production of heavy neutral lepton $N$ decaying from SM Higgs $h$. At the lepton colliders, there are two important production channels of SM Higgs, i.e., the Higgs-strahlung process $\ell^+\ell^-\to Zh$ and the $WW$ fusion process $\ell^+\ell^-\to \nu_\ell \bar{\nu}_\ell h$. It is well known that the cross section of the Higgs-strahlung process $\ell^+\ell^-\to Zh$ decreases at high energies as $\sim g^4/s$, where $\sqrt{s}$ is the center-of-mass energy. Meanwhile, the cross section of $WW$ fusion process $\ell^+\ell^-\to \nu_\ell \bar{\nu}_\ell h$ increases at high energies as $\sim g^4 \log(s/m_h^2)$ \cite{Kilian:1995tr}. At the TeV scale muon collider, the $WW$ fusion process is the dominant contribution \cite{Costantini:2020stv}, whose cross section can be derived in the high energy limit as \cite{Kilian:1995tr}
\begin{equation}
	\sigma(\mu^+\mu^-\to \nu_\mu \bar{\nu}_\mu h) \approx \frac{g^4}{256\pi^3} \frac{1}{v_0^2} \left(\log\frac{s}{m_h^2}-2\right).
\end{equation}
In this paper, we use {\bf Madgraph5\_aMC@NLO} \cite{Alwall:2014hca} to obtain more precise numerical results.

\begin{figure}
	\begin{center}
		\includegraphics[width=0.45\linewidth]{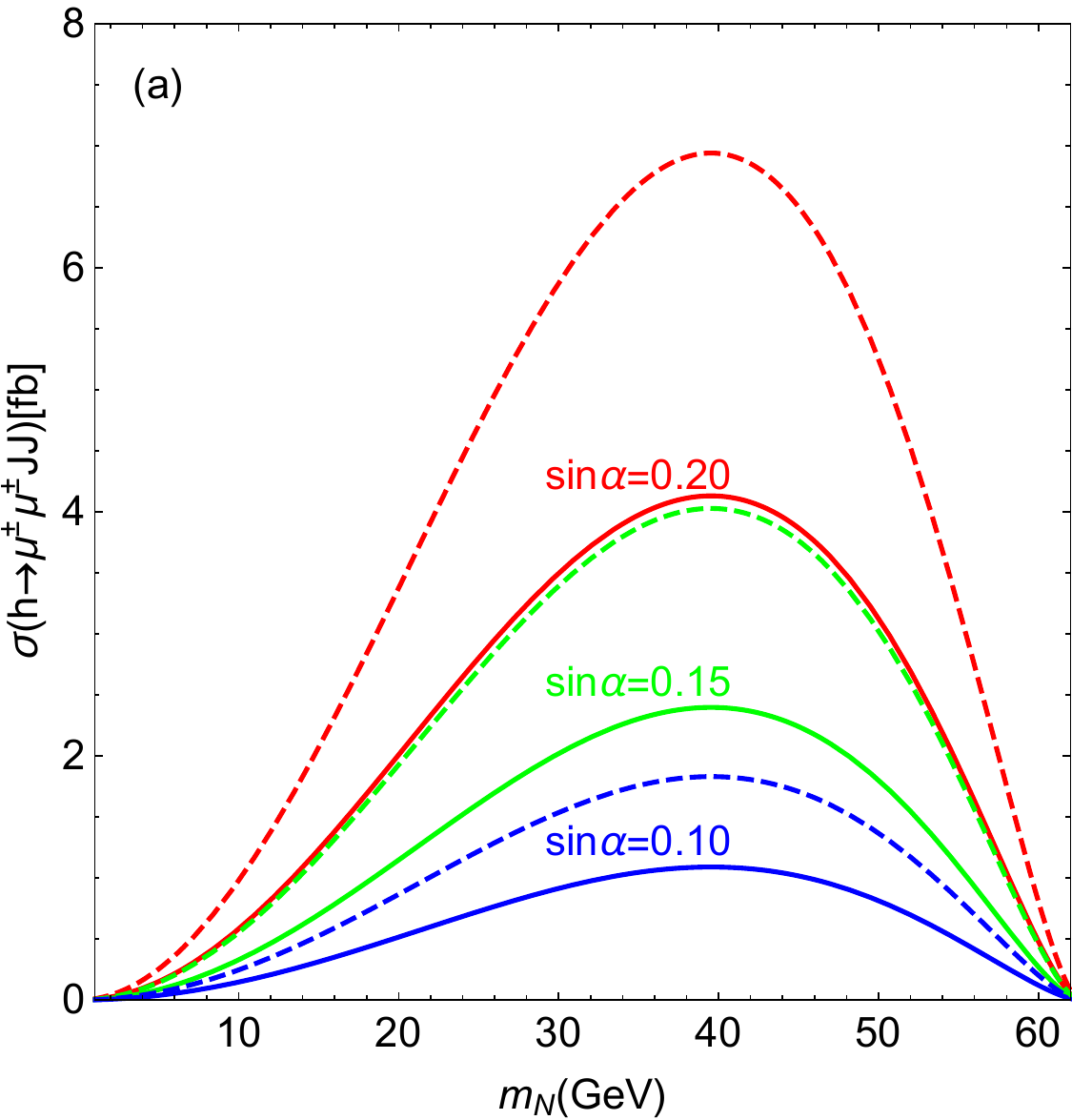}
		\includegraphics[width=0.45\linewidth]{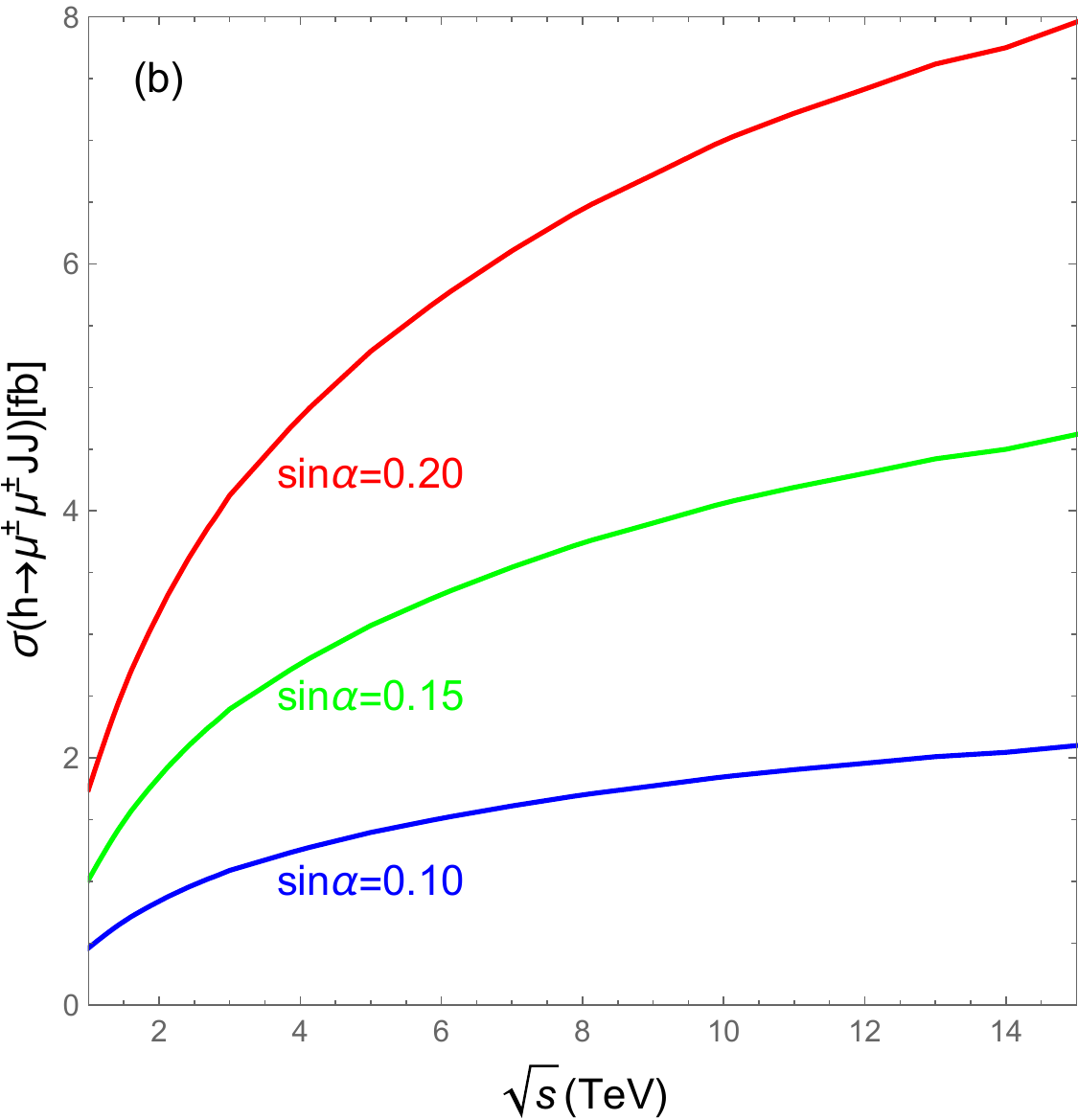}
	\end{center}
	\caption{Cross section of lepton number violation signature from SM Higgs decay as a function of heavy neutral lepton mass $m_N$ in panel (a) and the center-of-mass energy $\sqrt{s}$ in panel (b). The blue, green, and red lines are the results of the mixing  $\sin\alpha=0.2$, 0.15, and 0.1, respectively. In panel (a), the solid and dashed lines correspond to the results of 3 TeV and 10 TeV muon collider. In panel (b), we have fixed $m_N=40$ GeV.}
	\label{fig4}
\end{figure}

When the mass of heavy neutral lepton is small enough ($m_N<m_h/2$), the SM Higgs $h$ can decay into a couple of heavy neutral leptons through the mixing with heavy Higgs $H$. The full process of the lepton number violation signature of SM Higgs at the muon collider is
\begin{align}\label{Eqn:sig01}
	{\mu^+}{\mu^-}\rightarrow{\nu_\mu \bar{\nu}_\mu h}\rightarrow{NN \nu_\mu \bar{\nu}_\mu}\rightarrow{{\mu^\pm}jj}+{{\mu^\pm}jj}+\nu_\mu \bar{\nu}_\mu,
\end{align}
where we have assumed that the heavy neutral lepton $N$ couples exclusively with the muon flavor leptons for illustration, i.e., $|V_{\mu N}|\neq0$, and $|V_{e N}|=|V_{\tau N}|=0$. 
At the TeV scale muon collider, the dijets from heavy neutral lepton decays could be highly boosted and could merge into one fat-jet $J$. Hence, the signature is regarded as $\mu^\pm \mu^\pm JJ$.  For the electron mixing pattern, i.e., $|V_{e N}|\neq0$, and $|V_{\mu N}|=|V_{\tau N}|=0$, the cross section of the signal will be the same as the muon mixing pattern. However, the cross sections of the backgrounds involving electron are usually smaller than those involving muon \cite{Li:2023tbx}, so the electron mixing pattern is expected to be more promising than the muon mixing pattern at the muon collider. The muon mixing pattern in this paper represents a more conservative benchmark.

With narrow width approximation, the cross section of lepton number violation signature from SM Higgs decay can be calculated as
\begin{equation}
	\sigma(\mu^+\mu^-\xrightarrow{h} \mu^\pm\mu^\pm JJ) \simeq 	\sigma(\mu^+\mu^-\to \nu_\mu \bar{\nu}_\mu h) \times \text{BR}(h\to NN) \times \text{BR}(N\to \mu^\pm jj)^2/2.
\end{equation}

In Figure \ref{fig4}, we show the cross section for the $h\to\mu^\pm\mu^\pm JJ$ signature, which is sensitive to the mass of heavy neutral lepton $m_N$ and the mixing parameter $\sin\alpha$. With $m_N\simeq40$ GeV and $\sin\alpha =0.2$, we can achieve a cross section of about 4 fb at the 3 TeV muon collider. When $\sin\alpha<0.1$, the cross section is typically less than 1 fb at the 3 TeV muon collider. Increasing the collision energy will obtain a larger cross section. For instance, $\sigma(h\to \mu^\pm \mu^\pm JJ)$ could reach about 7 fb at the 10 TeV muon collider for $\sin\alpha=0.2$. With also higher integrated luminosity \cite{Accettura:2023ked}, the 10 TeV muon collider seems more promising than the 3 TeV one. 

In this work, we consider the main SM backgrounds as
\begin{align}
	{\mu^+}{\mu^-}&\rightarrow {\mu^+}{\mu^-}WW,\\ 
	{\mu^+}{\mu^-}&\rightarrow{\mu^+}{\mu^-}WWZ,\\
	{\mu^+}{\mu^-}&\rightarrow \mu^{\pm}{\nu}WWW.
\end{align}
The contributions of ${\mu^+}{\mu^-}\rightarrow {\mu^+}{\mu^-}WW$ and ${\mu^+}{\mu^-}\rightarrow {\mu^+}{\mu^-}WWZ$ are from lepton charge misidentified, which are significantly suppressed by the misidentification rate $0.1\%$ \cite{Liu:2021akf,ATLAS:2019jvq}. Backgrounds from the ${\mu^+}{\mu^-}\rightarrow \mu^{\pm}{\nu}WWW$ process is dominant by the vector boson fusion, where the two same-sign $W$ bosons decay leptonically. Other backgrounds from $\mu^+\mu^+\to \mu^+\mu^- jj$, ${\mu^+}{\mu^-}\rightarrow {W^\pm}{W^\pm}jj$ and ${\mu^+}{\mu^-}\rightarrow t \bar{t}{W^\pm}{\mu^\mp}v$ are not considered, as they can be greatly reduced by proper cuts \cite{He:2024dwh}.  For example, after the selection cuts in Table \ref{Tab02}, the total cross section of these additional backgrounds is $8.4\times10^{-5}$ fb at the 3 TeV muon collider, which is over two orders of magnitude smaller than the dominant backgrounds. So including these additional backgrounds has a negligible impact on the final results. 

In the simulation, we use {\bf FeynRules}  package \cite{Alloul:2013bka} to implement the scalar singlet extension of the type-I seesaw model. We then use {\bf Madgraph5\_aMC@NLO} \cite{Alwall:2014hca} to generate the leading order signal and background events at the parton level. After this, {\bf Pythia8} \cite{Sjostrand:2014zea} is used to do parton showering and hadronization, and the detector simulation is performed by {\bf Delphes3} \cite{deFavereau:2013fsa} with the corresponding muon collider card. The fat-jets are reconstructed by using the Valencia algorithm \cite{Boronat:2014hva} with $R = 1.2$.

Pre-selection cuts on the transverse momentum of fat-jets, pseudorapidity of the muon and the fat-jets are applied as following
\begin{align}
	|\eta(\mu)|<2.5, P_T(J)>50~{\rm{GeV}}, |\eta(J)|<2.5,
\end{align}
 which correspond to the default set of {\bf Delphes3}.

\begin{figure}
	\begin{center}
		\includegraphics[width=0.45\linewidth]{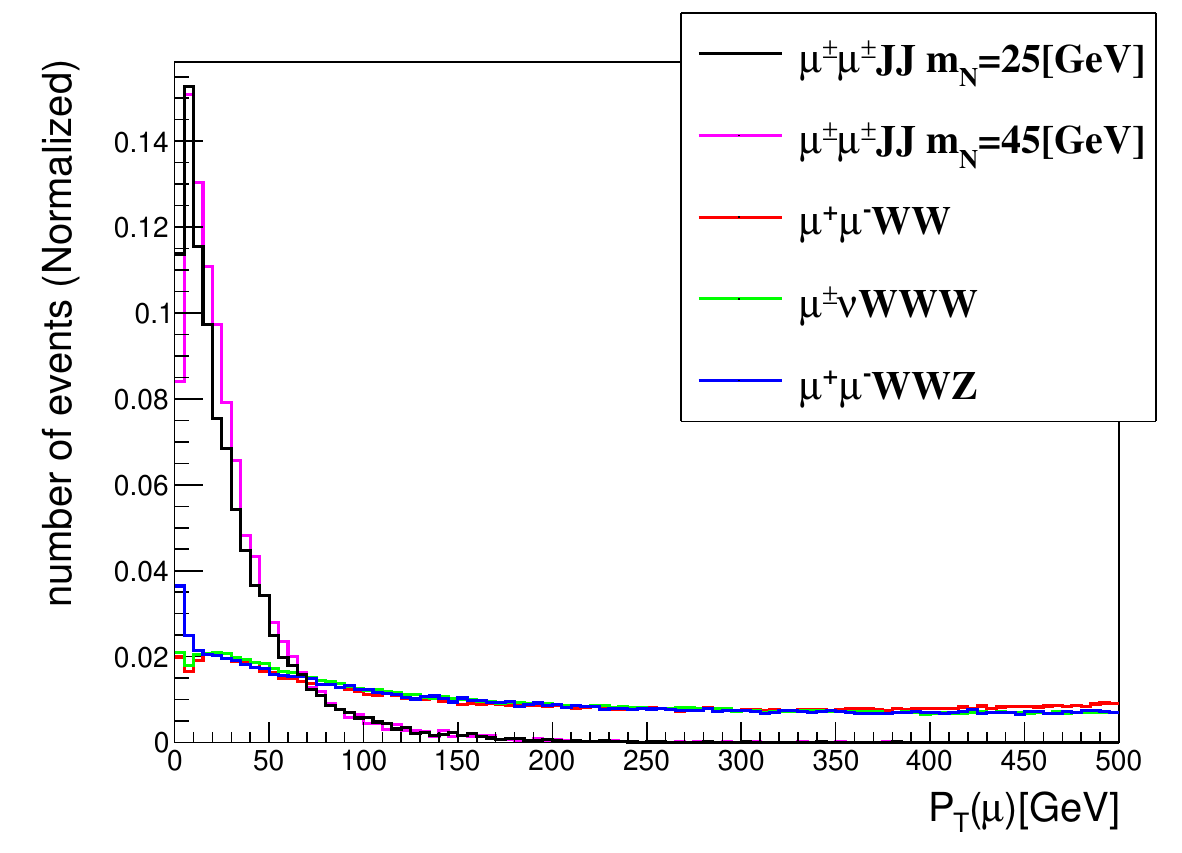}
		\includegraphics[width=0.45\linewidth]{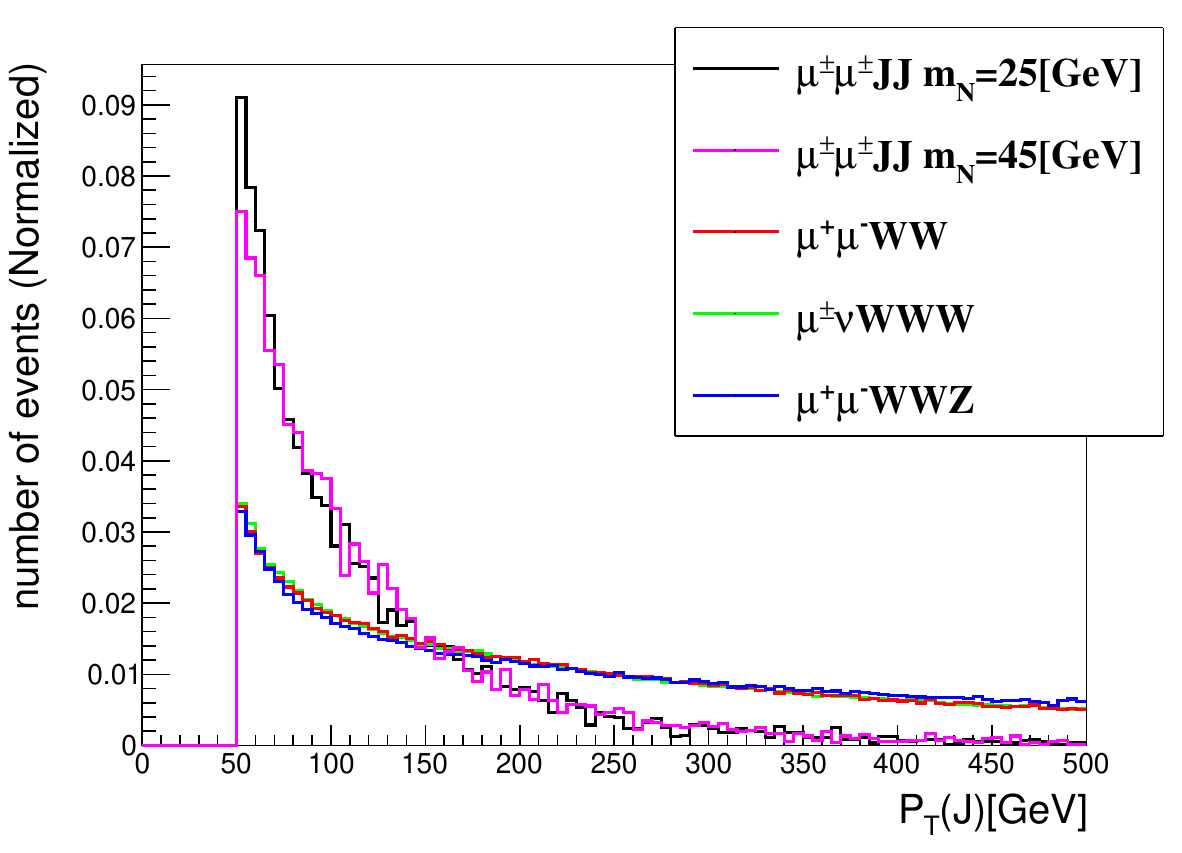}
		\includegraphics[width=0.45\linewidth]{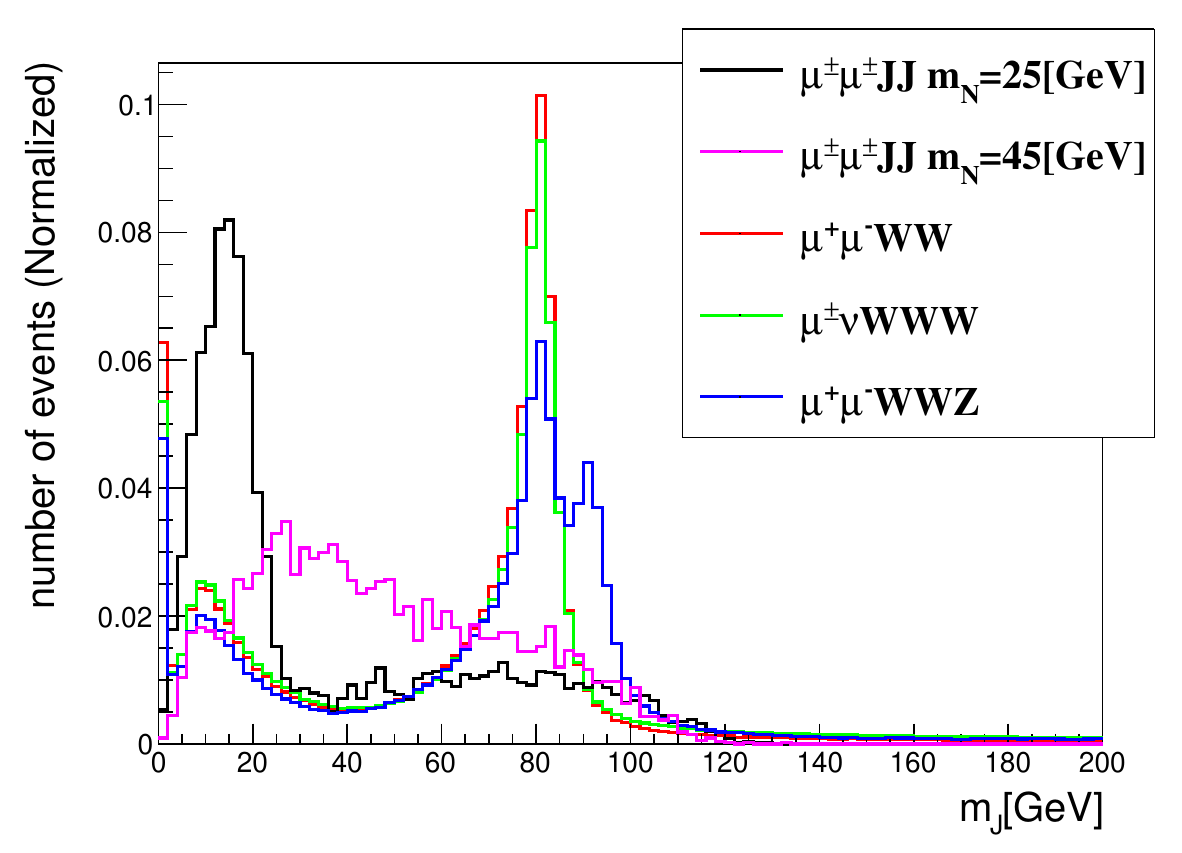}
		\includegraphics[width=0.45\linewidth]{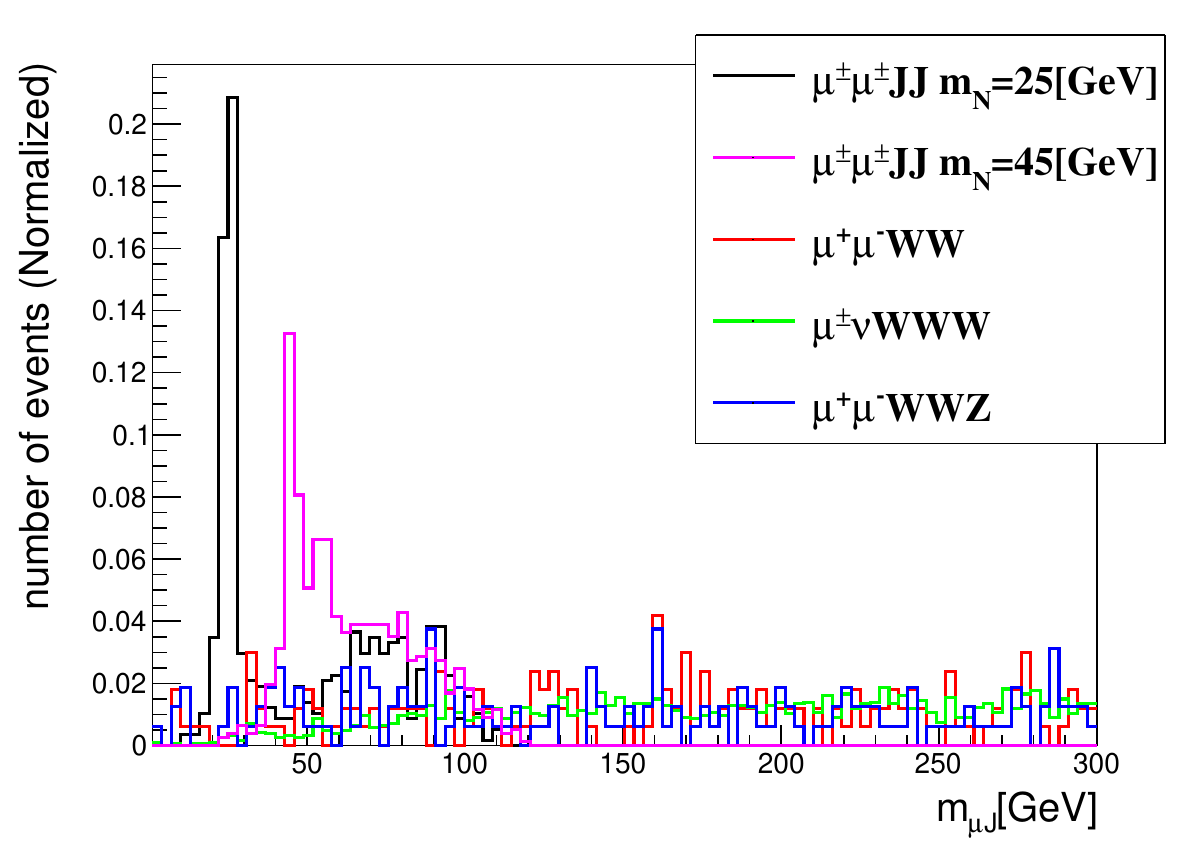}
	\end{center}
	\caption{Normalized distributions of muon transverse momentum $P_T(\mu)$ (up-left panel), fat-jet transverse momentum $P_T(J)$ (up-right panel), invariant mass of fat-jet $m_{J}$ (down-left panel), and invariant mass of muon and fat-jet $m_{\mu{J}}$ (down-right panel)  for the signal and corresponding backgrounds at the 3 TeV muon collider. We choose two benchmark points with $m_N=25~\rm{GeV}$ and $45~\rm{GeV}$. }
	\label{fig5}
\end{figure}

In Figure \ref{fig5}, we show the distributions of some variables for the signal and backgrounds at the 3 TeV muon collider after the pre-selection cuts, where $m_N=25$GeV and 45GeV with $\sin\alpha=0.2$ are chosen as the benchmark points. Distributions of these variables at the 10 TeV muon collider are similar to the 3 TeV muon collider, hence they will not be shown in this paper.  The distributions are normalized by using number of event in each bin divided by the total events. The normalized distributions are used to compare the difference between the signal and the backgrounds at the same level, based on which further selection cuts are applied.

After the simulation, several cuts are further applied to suppress the SM backgrounds. To obtain the lepton number violation signature $\mu^\pm\mu^\pm JJ$, we first require exactly two same-sign muons in the final states. According to the distribution of $P_T(\mu)$ in Figure \ref{fig5}, the muons from  heavy neutral lepton decays typically have a transverse momentum smaller than 100 GeV, while those from SM backgrounds could be much more energetic. So we select events with $P_T(\mu)$ in the range of 20 to 100 GeV. The cuts on muons are summarized as
\begin{align}\label{cut01}
	N_\mu^{\pm}=2, 20~{\rm{GeV}}<P_T(\mu)<100~\rm{GeV}.
\end{align}

In Table \ref{Tab01}, we show the cut flow for the $h\to {\mu^\pm}{\mu^\pm}JJ$ signature  of the benchmark point with $m_N=45$~GeV, $\sin\alpha=0.2$ and the dominant SM backgrounds, where the results in the brackets correspond to the 10 TeV scenario. It is obvious that the selection cuts on the same-sign dimuon are enough to suppress the backgrounds smaller than the signal. The dominant background is from the $\mu^\pm\nu WWW$ process after the muon cuts. At the 3~TeV muon collider,  the total cross section of the backgrounds is at the order of $\mathcal{O}(10^{-2})$ fb, which is an order of magnitude smaller than the signal. Increasing the collision energy to 10 TeV, the muons from gauge boson decays in the backgrounds are more energetic, which results in much smaller efficiency to pass the muon cuts $P_T(\mu)<100$ GeV in Equation \eqref{cut01}.

We also find that the acceptance efficiency of the signal $\sim0.15$ is  relatively small compared to the efficiency of the signal from heavy Higgs. This is because the $\mathcal{O}(10)$~GeV scale heavy neutral leptons at the TeV muon collider are also boosted, which results in the muons tending to be non-isolated from the jets. Using the substructure based variables as lepton sub-jet fraction (LSF) and lepton mass drop (LMD) \cite{Brust:2014gia}, such events with non-isolated muons can be further identified to enhance the significance of this signature, which will be considered in future studies.

To make sure the same-sign dimuon come from the heavy neutral lepton, cuts on fat-jets are required. It is clear in Figure \ref{fig5}  that there are no peak structures around $m_W$ in the distribution of invariant mass of fat-jet $m_J$ for the signal, which conforms to the fact that these fat-jets are from the off-shell $W$ boson decay.  The distributions of $m_J$ for the signal depend on the value of $m_N$. For heavier $m_N$, e.g., $m_N=45$ GeV, the distribution of $m_J$ will have a larger overlap with the background. We do not apply any cuts on $m_J$ in this analysis for simplicity. So the results in Table \ref{Tab01} might be improved by adopting $m_N$ dependent cut on $m_J$, e.g., $m_J<30$ GeV when $m_N=25$ GeV. 
As the same-sign dimuon cuts are already enough to suppress the SM backgrounds much smaller than the signal, we only require at least one fat-jet $J$ in the final states to keep as many signal events as possible. Based on the distribution of $P_T(J)$ in Figure~\ref{fig5}, we also require the transverse momentum of fat-jets to be less than 200 GeV. The fat-jet cuts are 
\begin{align}\label{cut02}
	N_J\geq1,50{\rm{GeV}}<P_T(J)<200\rm{GeV}.
\end{align}

\begin{table}
	\begin{center}
		\begin{tabular}{c | c | c | c | c} 
			\hline
			\hline
			$\sigma(\text{fb})$&${\mu^\pm}{\mu^\pm}JJ$&${\mu^+}{\mu^-}WW$&$\mu^\pm{\nu}WWW$&$\mu^+\mu^-WWZ$ \\
			\hline
			pre-selection & 3.9(13) &28(13) &5.8(11) &0.84(1.3) \\
			\hline
			muon cuts  &0.63(1.8) & $3.0\times10^{-3}$($2.3\times10^{-4})$ & $2.1\times10^{-2}$($3.1\times10^{-3}$) & $2.4\times10^{-4}$($4.8\times10^{-5}$)\\
			\hline
			fat-jet cuts &0.34(0.99)  & $9.5\times10^{-4}$($5.2\times10^{-5}$)  & $1.1\times10^{-2}$($7.2\times10^{-4}$) & $1.1\times10^{-4}$($1.0\times10^{-5}$) \\
			\hline
			$m_{\mu{J}}$ cut  &0.13(0.36)  & $5.6\times10^{-5}$($2.6\times10^{-5}$) & $4.1\times10^{-4}$($2.2\times10^{-5}$) &$1.7\times10^{-6}$($2.4\times10^{-6}$)\\
			\hline
			\hline
			Significance & 34(238) & \multicolumn{2}{|c|}{Total Background} & $4.7\times10^{-4}$($5.0\times10^{-5}$) \\
			\hline
		\end{tabular}
	\end{center}
	\caption{Cut flow table for the signal and three dominant backgrounds process with $m_N=45~\rm{GeV}$ and $\sin\alpha=0.2$ at the 3TeV (10TeV) muon collider. The significance is calculated with an integrated luminosity of 1 (10) ab$^{-1}$ for the 3TeV (10TeV) muon collider. }
	\label{Tab01}
\end{table}

Then the mass of heavy neutral lepton $m_N$ can be reconstructed by the invariant mass $m_{\mu J}$, which is further used to select events in the mass range of
\begin{align}\label{cut03}
	0.8m_N<m_{\mu{J}}<1.2m_N.
\end{align}

After all the selection cuts, the total cross section of the SM background is $4.7\times10^{-4}$ fb at the 3 TeV muon collider and $4.8\times10^{-5}$ fb at the 10 TeV muon collider. Therefore, with an integrated luminosity of 1(10) ab$^{-1}$, there would be 0.47(0.48) background events at the 3(10) TeV muon collider. For the benchmark point with $m_N=$45 GeV and $\sin\alpha=0.2$, we can get about 130(3600) signal events with 1(10)~ab$^{-1}$, which corresponds to a significance of $S=34(238)$. The significance is calculated by using the algorithm proposed in Ref~\cite{Cowan:2010js} as
\begin{align}
	S=\sqrt{2\left[(N_S+N_B)\log\left(1+\frac{N_S}{N_B}\right)-N_S\right]},
\end{align}
where $N_S$ and $N_B$ are the event number of signal and backgrounds, respectively. To reach the $5\sigma$ discovery limit for the benchmark point, the required luminosity is about  21(4.4) fb$^{-1}$ at the 3(10) TeV muon collider. Therefore, the lepton number violation SM Higgs decay signature is quite promising at the TeV scale muon collider.
 We then estimate the precision for the signature by using $\sqrt{N_S+N_B}/N_S$ \cite{Forslund:2022xjq}. For the benchmark point in Table \ref{Tab01}, we obtain a precision of 8.8\% at the 3 TeV and 1.7\% at the 10 TeV muon collider after all selection cuts.

\begin{figure}
	\begin{center}
    \includegraphics[width=0.45\linewidth]{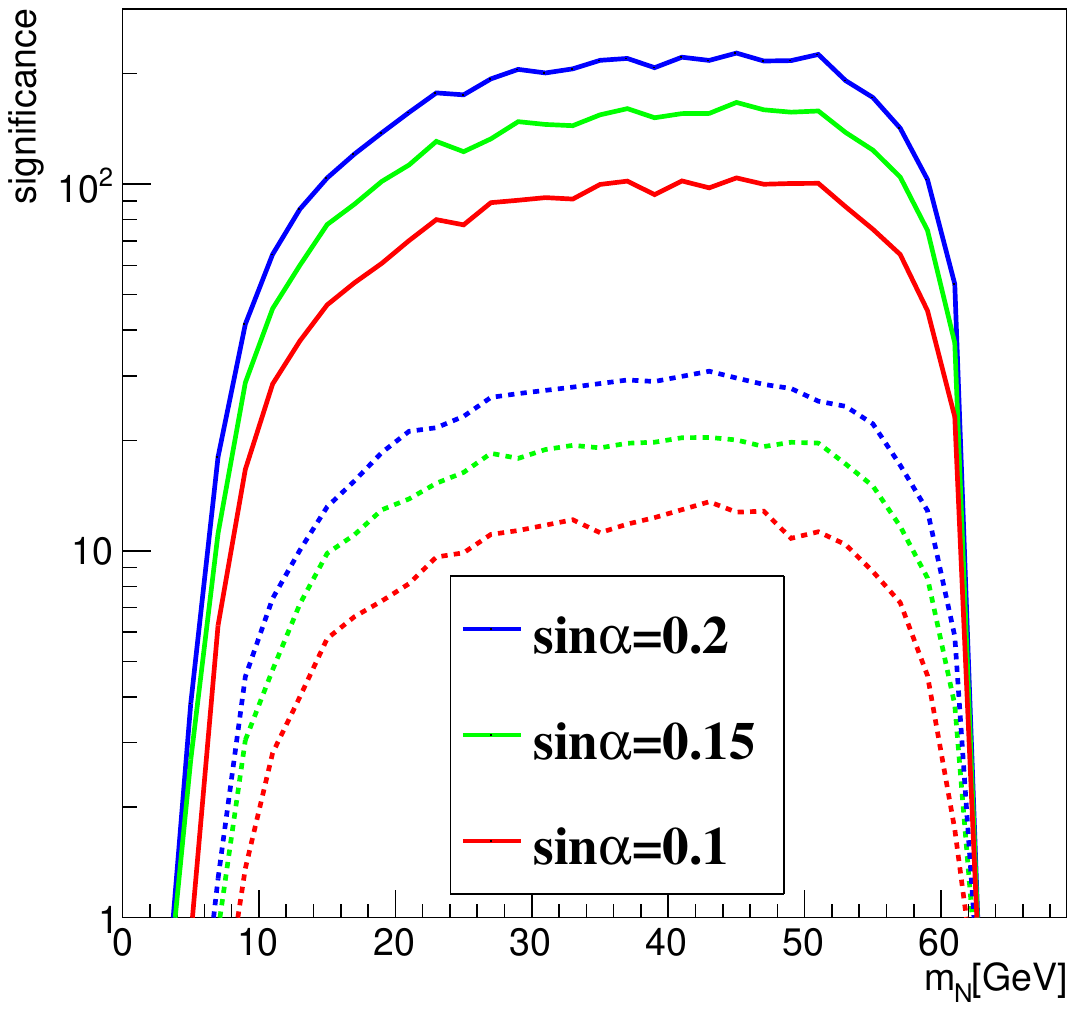}
    \includegraphics[width=0.45\linewidth]{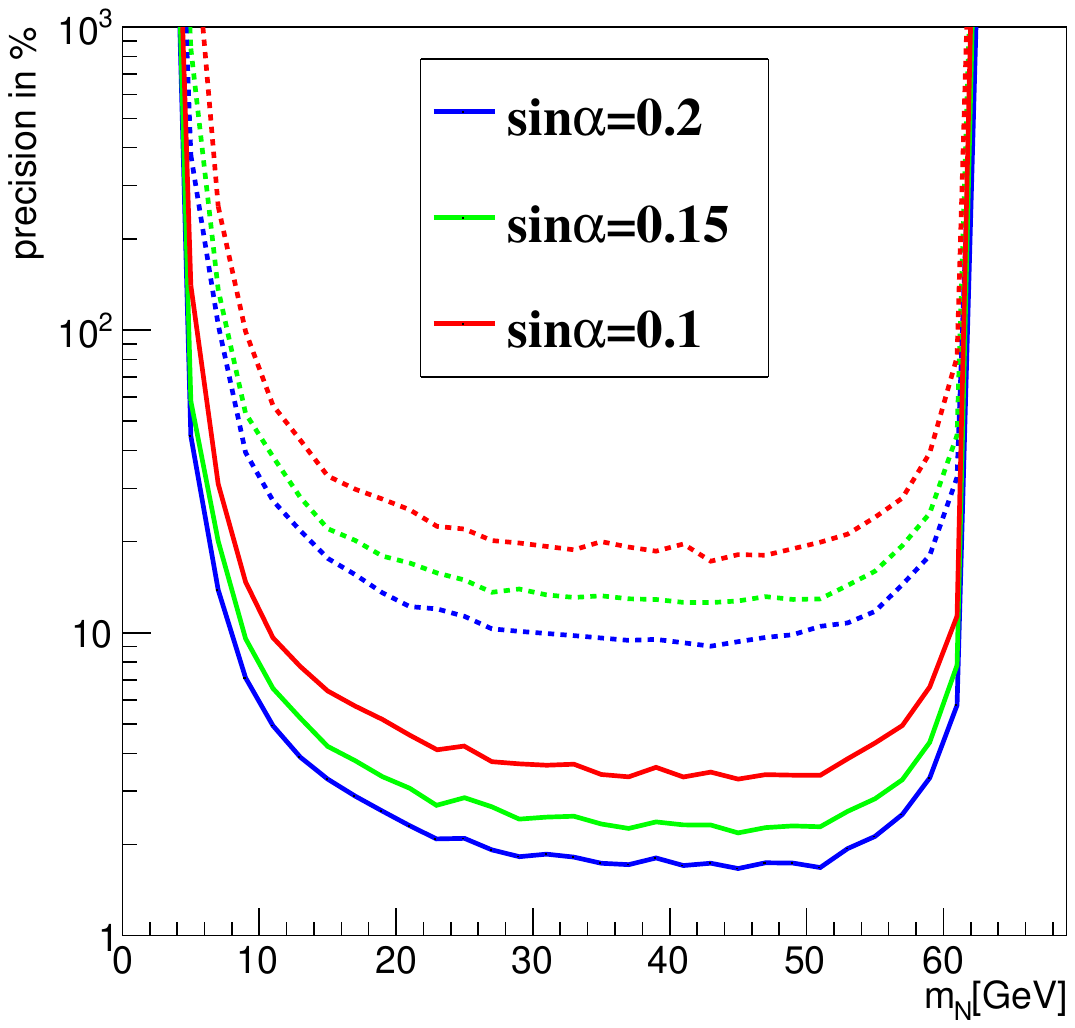}
    \includegraphics[width=0.45\linewidth]{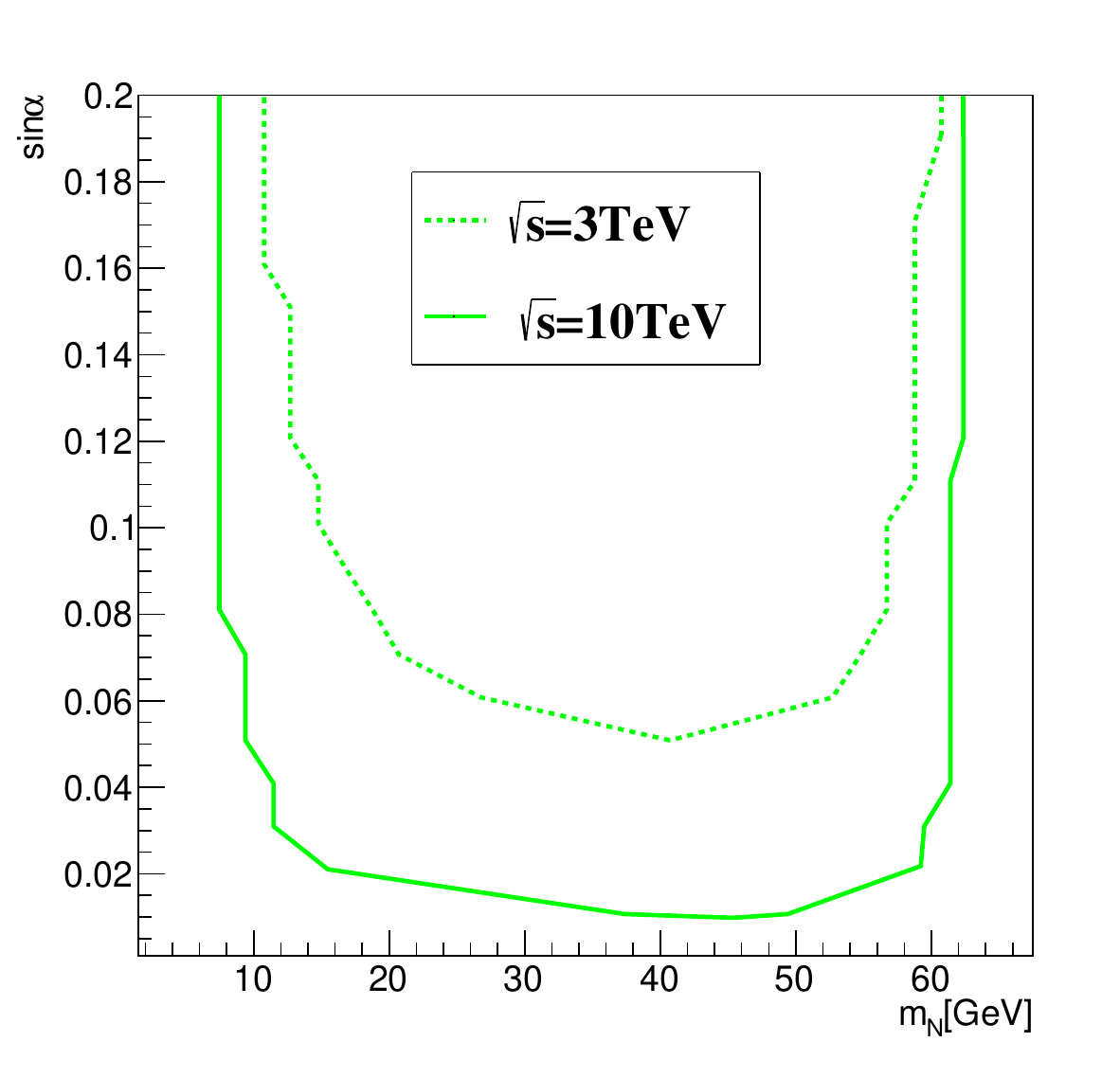}
	\end{center}
	\caption{Significance (upper-left panel) and precision (upper-right panel) for the lepton number violation SM Higgs decay $h\to \mu^\pm \mu^\pm JJ$. The blue, green, and red lines are the results with $\sin{\alpha}=0.2,0.15$, and 0.1 respectively. The dashed (solid) lines are for 3 (10) TeV muon collider with an integrated luminosity of $L=1 (10)~\text{ab}^{-1}$. Lower panel: The $5\sigma$ sensitivity reach of the $h\to\mu^{\pm}\mu^{\pm}JJ$ signature at the 3 TeV (dashed line) and 10 TeV (solid line) muon collider with an integrated luminosity of $L=1~\text{ab}^{-1}$ and $L=10~\text{ab}^{-1}$, respectively. }
	\label{fig6}
\end{figure}

The signature of lepton number violation SM Higgs decay $h\to \mu^\pm \mu^\pm JJ$ depends on the mass of heavy neutral lepton $m_N$ and the Higgs mixing parameter $\sin\alpha$. We then perform a cut-based analysis to explore the sensitive region. In the upper-left panel of Figure  \ref{fig6}, we show the significance of the $h\to{\mu^\pm}{\mu^\pm}JJ$ signature by varying the mass of heavy neutral leptons $m_N$ for three values of mixing angle $\sin\alpha=0.2,0.15$ and 0.1 at the muon collider. With an integrated luminosity of $L=1(10)~\text{ab}^{-1}$, the $h\to{\mu^\pm}{\mu^\pm}JJ$ signature can be discovered in the mass range of $10(5)~\text{GeV}<m_N<60~\text{GeV}$ for $\sin\alpha=0.2$ at the 3 (10) TeV muon collider.  For $m_N$ below 10 GeV, the final state muon from boosted $N$ decay can hardly be separated from the jets, which leads to the quickly decrease of the significance.  The largest significance could reach about 30(230) when $m_N\simeq45$ GeV for the collision energy of 3 (10) TeV.  A smaller mixing angle will lead to a smaller discovery mass range. The 3 TeV muon collider is promising to cover the mass range of $16~\text{GeV}<m_N<56~\text{GeV}$  when $\sin\alpha>0.1$.  The precision results are shown in the upper-right panel of Figure \ref{fig6}. Typically, the 3 TeV muon collider could have a precision of $\mathcal{O}(10)$\% for $\sin\alpha\gtrsim0.1$. And the 10 TeV muon collider could reach the precision of $\mathcal{O}(2)$\% for $\sin\alpha\simeq0.2$ with 25 GeV $\lesssim m_N\lesssim$ 55 GeV.

In the lower panel of Figure \ref{fig6}, we depict the $5\sigma$ discovery reach of the $h\to\mu^{\pm}\mu^{\pm}JJ$ signature at the 3(10) TeV muon collider with an integrated luminosity of $L=1(10)~\text{ab}^{-1}$. When $m_N\simeq45~\rm{GeV}$, the limit of the mixing parameter $\sin\alpha$ can be as small as $\sin\alpha=0.05(0.009)$ at the 3(10) TeV muon collider, which corresponds to the branching ratio of $h\to NN$ as $4.1\times10^{-3}(1.3\times10^{-4})$.

%%%%%%%%%%%%%%%%%%%%%%%%%%%%%%%%%%%%

\section{Signature from Heavy Higgs}\label{Sec:HH}

Through mixing with SM Higgs $h$, the dominant production channel of heavy Higgs $H$ at the TeV scale muon collider is also via vector boson fusion process $\mu^+\mu^-\to \nu_\mu \bar{\nu}_\mu H$. In the high energy limit, the corresponding cross section can be estimated as \cite{Buttazzo:2018qqp}
\begin{equation}\label{Eqn:csH}
	\sigma(\mu^+\mu^-\to \nu_\mu \bar{\nu}_\mu H) \approx \frac{g^4 \sin^2\alpha}{256\pi^3} \frac{1}{v_0^2} \left(\log\frac{s}{m_H^2}-2\right).
\end{equation}
The numerical results are also obtained with {\bf Madgraph5\_aMC@NLO} \cite{Alwall:2014hca} in the following discussions. When kinematically allowed $m_N<m_H/2$, the heavy Higgs could decay into the heavy neutral lepton pair. The lepton number violation signature of heavy Higgs at muon collider is
\begin{align}
	{\mu^+}{\mu^-}\rightarrow{\nu_\mu \bar{\nu}_\mu H}\rightarrow{NN\nu_\mu \bar{\nu}_\mu}\rightarrow{{\mu^\pm}jj}+{{\mu^\pm}jj}+\nu_\mu \bar{\nu}_\mu.
\end{align}
Similar to the discussion in Section \ref{Sec:SM}, we also assume that the heavy neutral lepton $N$ couples exclusively with the muon flavor leptons. The dijets from heavy neutral lepton decays are reconstructed as one fat-jet $J$. The lepton number violation signature $H\to \mu^\pm \mu^\pm JJ$ has the same SM backgrounds as in Section \ref{Sec:SM}, i.e., $\mu^+\mu^-\to \mu^+\mu^-WW, \mu^+\mu^- WWZ, \mu^\pm \nu WWW$.

It should be noted that when the condition $m_H<2m_W$ is satisfied, we can still have the lepton number violation signature $H\to \mu^\pm\mu^\pm JJ$ with $m_N<m_H/2$ \cite{Nemevsek:2016enw}, but without the fat-jet mass satisfying $m_J\sim m_W$. In this relatively light scalar singlet $H$ scenario, the decay $H\to \mu^\pm\mu^\pm JJ$  is similar to the SM Higgs decay $h\to \mu^\pm\mu^\pm JJ$, which will also contribute to the lepton number violation signature in Section \ref{Sec:SM}. Therefore, the significance of the $\mu^\pm\mu^\pm JJ$ signature in Section \ref{Sec:SM} could be enhanced in this light scalar singlet scenario.

For the sake of difference, we consider that the heavy neutral lepton is heavier than the $W$ boson. So the SM Higgs decay $h\to NN$ is forbidden in this scenario. Different from the signature $h\to \mu^\pm \mu^\pm JJ$ with fat-jets from the off-shell $W$ decay, the fat-jets in the signature $H\to \mu^\pm \mu^\pm JJ$ are from the on-shell $W$ decay when $m_N>m_W$. Hence, we expected the invariant mass of fat-jet satisfying $m_J\sim m_W$ in this scenario, which can be used to distinguish the $H\to \mu^\pm \mu^\pm JJ$ signature from the $h\to \mu^\pm \mu^\pm JJ$ signal.

\begin{figure}
	\begin{center}
		\includegraphics[width=0.48\linewidth]{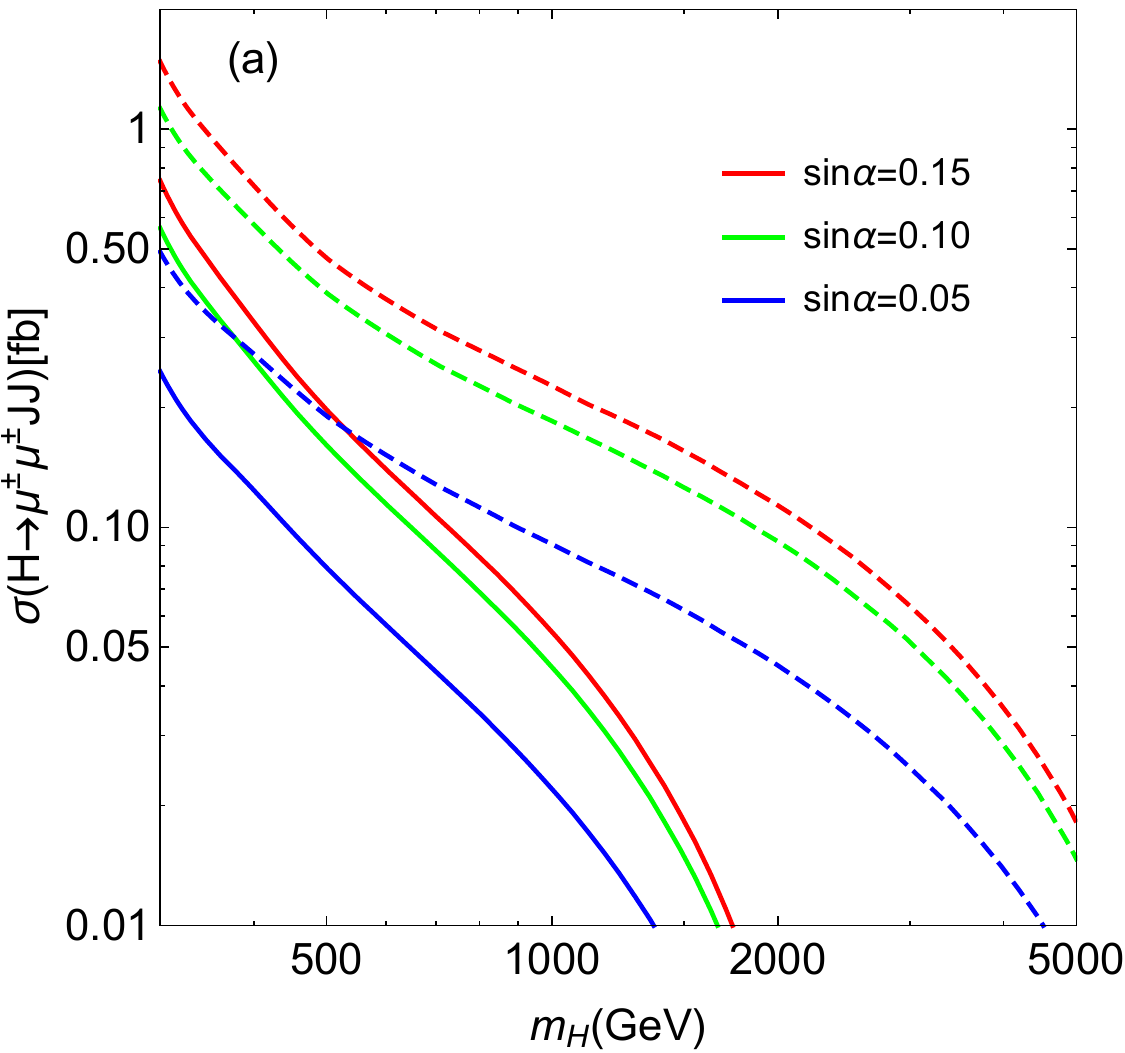}
		\includegraphics[width=0.45\linewidth]{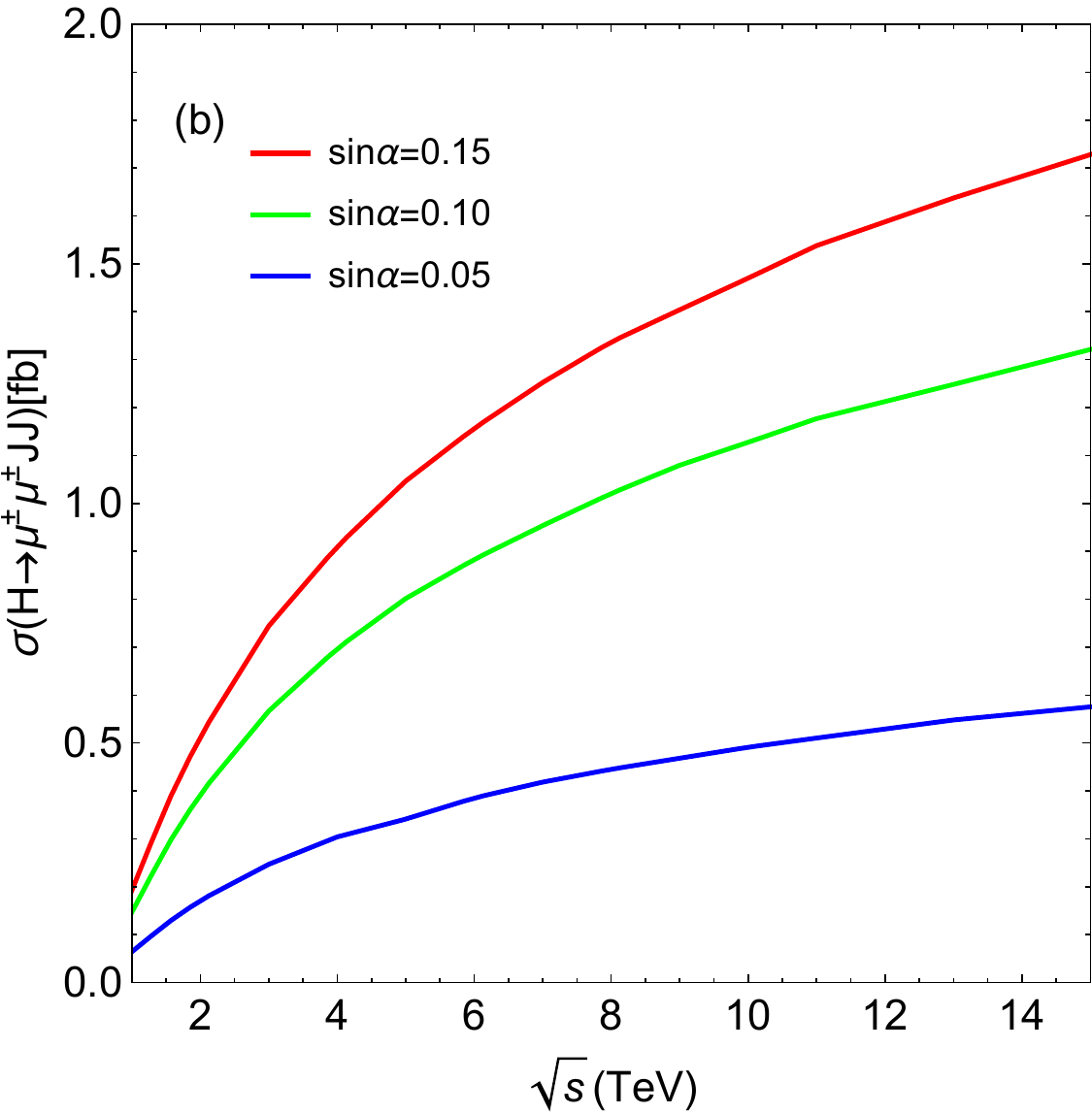}
		\includegraphics[width=0.45\linewidth]{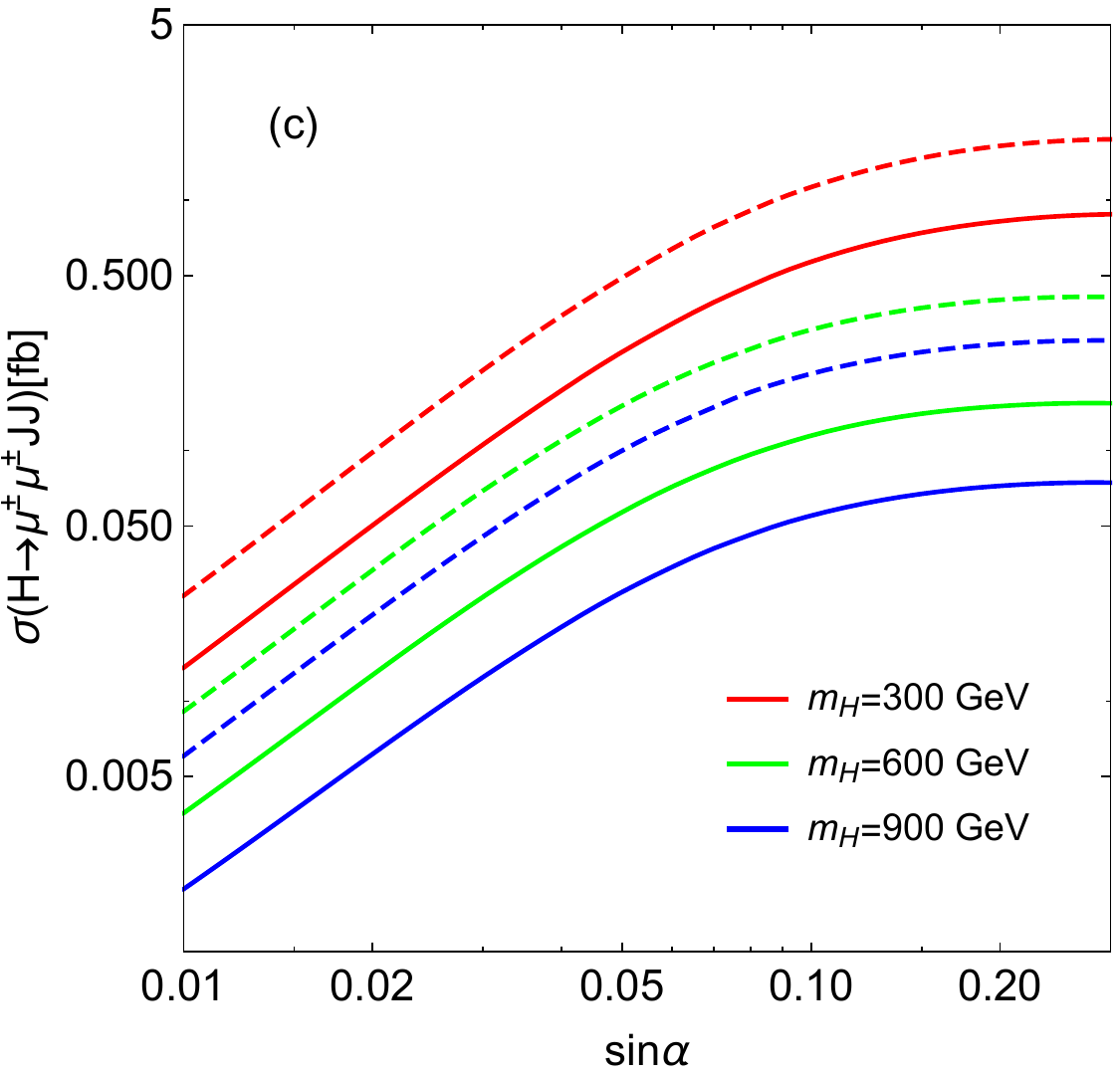}
		\includegraphics[width=0.45\linewidth]{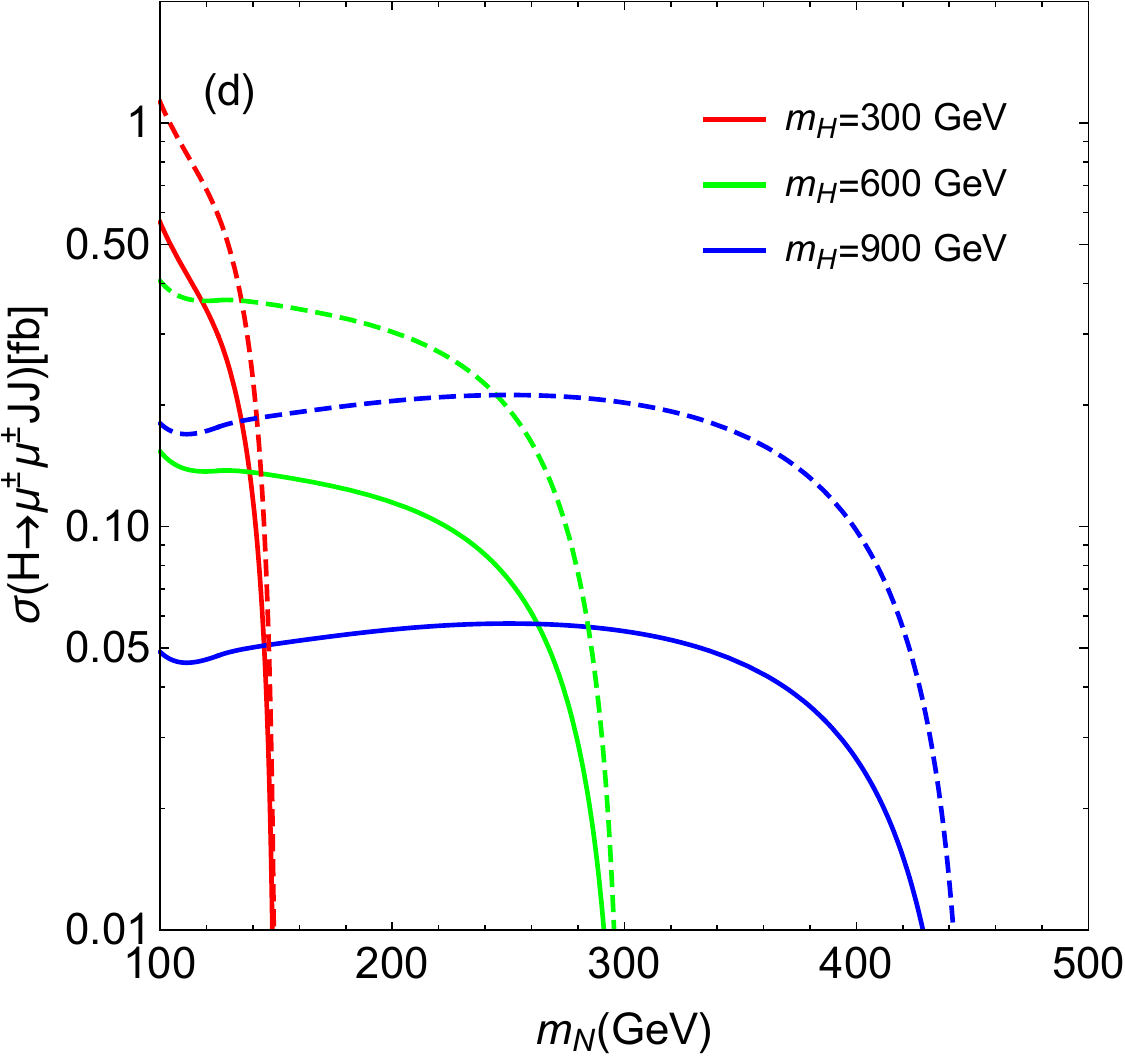}
	\end{center}
	\caption{Cross section of lepton number violation heavy Higgs decay. In panels (a), (b) and (c), we have fixed the mass relation as $m_N=m_H/3$. In panel (b), we further set $m_H=300$ GeV. In panel (d), the Higgs mixing parameter is fixed as $\sin\alpha=0.1$. In panels (a), (c), and (d), the solid lines are the results of the 3 TeV muon collider, while the dashed lines are the results of the 10 TeV muon collider.}
	\label{fig7}
\end{figure}

The theoretical cross section of the lepton number violation signature from heavy Higgs decay can be calculated as
\begin{equation}
	\sigma(\mu^+\mu^-\xrightarrow{H} \mu^\pm\mu^\pm JJ) \simeq 	\sigma(\mu^+\mu^-\to \nu_\mu \bar{\nu}_\mu H) \times \text{BR}(H\to NN) \times \text{BR}(N\to \mu^\pm jj)^2/2,
\end{equation}
which depends on the heavy Higgs mass $m_H$, the collision energy $\sqrt{s}$, the Higgs mixing parameter $\sin\alpha$ and the heavy neutral lepton mass $m_N$. 

In Figure \ref{fig7}, we show the numerical results of $\sigma(H\to \mu^\pm \mu^\pm JJ)$ at the muon collider. In panel (a),  the cross section at 3 TeV and 10 TeV muon collider as a function of $m_H$ are shown, where we set $m_N=m_H/3$ for illustration. We should mention that for $\sin\alpha=0.15$, the current direct searches of heavy Higgs boson could exclude the mass region of 300 GeV$\lesssim m_H\lesssim500$ GeV, meanwhile, all the mass range is allowed for $\sin\alpha<0.1$ \cite{Lane:2024vur}.  The cross section decreases as the heavy Higgs mass increases. Typically, we have $\sigma(H\to \mu^\pm \mu^\pm JJ)>0.01$ fb when $m_H\lesssim1.5 (5)$ TeV at the 3 (10) TeV muon collider. Hence, increasing the collision energy of the muon collider to 10 TeV is able to probe heavy Higgs above the TeV scale. In panel (b) of Figure \ref{fig7}, we show the cross section of $H\to \mu^\pm \mu^\pm JJ$ as a function of the center-of-mass energy $\sqrt{s}$, which clearly depicts the logarithmic dependence of collision energy as in Equation \ref{Eqn:csH}.

Another important parameter that alters the signal cross section is the Higgs mixing angle $\alpha$. In panel (c) of Figure \ref{fig7}, we show the cross section of $H\to \mu^\pm \mu^\pm JJ$ by varying $\sin\alpha$. Although the production cross section of heavy Higgs $\sigma(\mu^+\mu^-\to \nu_\mu \bar{\nu}_\mu H)$ decreases as $\sin\alpha$ becomes smaller according to Equation \ref{Eqn:csH}, the branching ratio of $H\to NN$ increases when $\sin\alpha$ decreases as shown in Figure \ref{fig2}. Therefore, the signal cross section $\sigma(H\to \mu^\pm\mu^\pm JJ)$ is nearly a constant when $\sin\alpha>0.1$. On the other hand, for $\sin\alpha<0.05$, the branching ratio  BR$(H\to NN)$ is already close to one, so the signal cross section becomes proportional to $\sin^2\alpha$, and decreases quickly as $\sin\alpha$ becomes smaller.

In panel (d) of Figure \ref{fig7}, the impacts of heavy neutral lepton mass $m_N$ on the signal cross section are shown, where we have fixed $m_H=$ 300 GeV, 600 GeV and 900 GeV for illustration.  We find that when $m_N\lesssim m_H/3$, the signal cross section $\sigma(H\to \mu^\pm\mu^\pm JJ)$ is approximately a constant. This is because the branching ratio of $H\to NN$ increases (due to larger Yukawa coupling $y_N=\sqrt{2} m_N/v_s$) while the branching ratio of $N\to \mu^\pm W^\mp$ decreases (as shown in Figure \ref{fig3} ) when $m_N$ becomes larger for certain $m_H$. Close to the threshold $m_N\sim m_H/2$, the signal cross section will be suppressed by the phase space of $H\to NN$ decay.

\begin{figure}
	\begin{center}
		\includegraphics[width=0.45\linewidth]{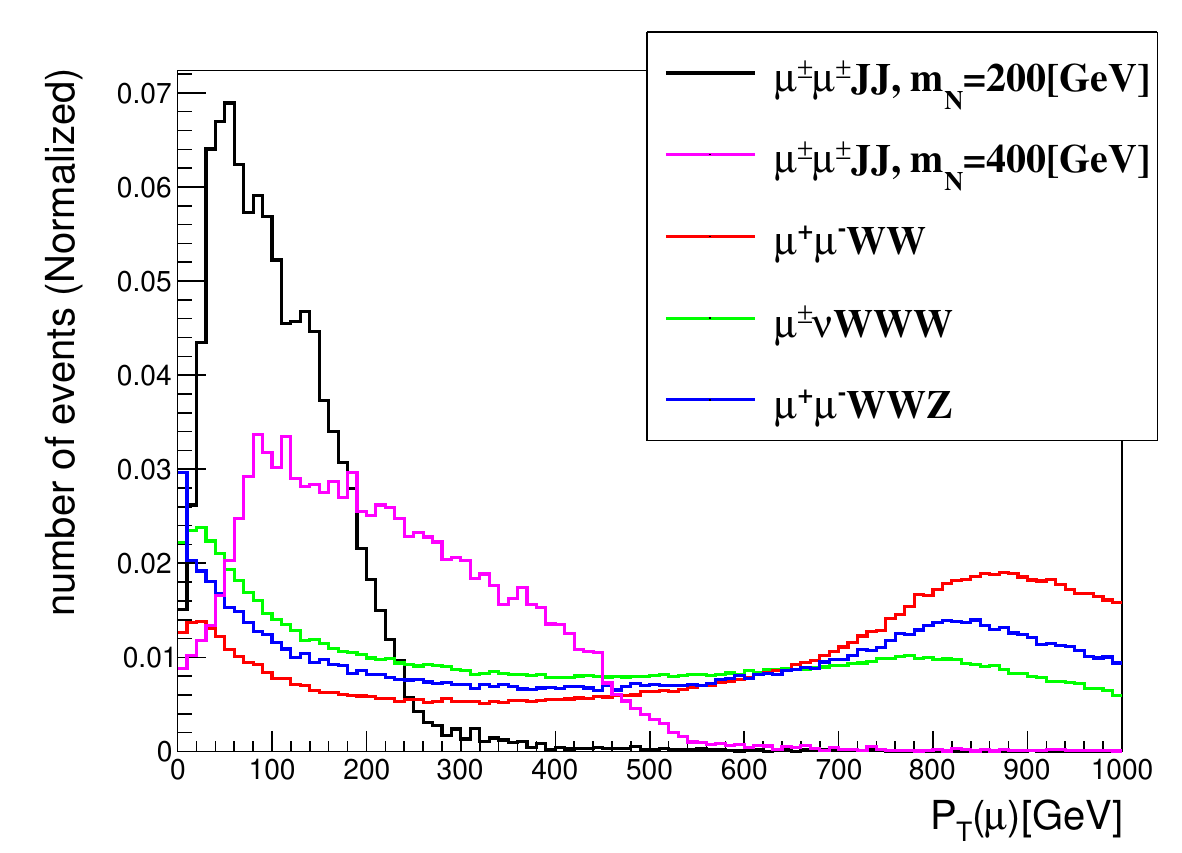}
		\includegraphics[width=0.45\linewidth]{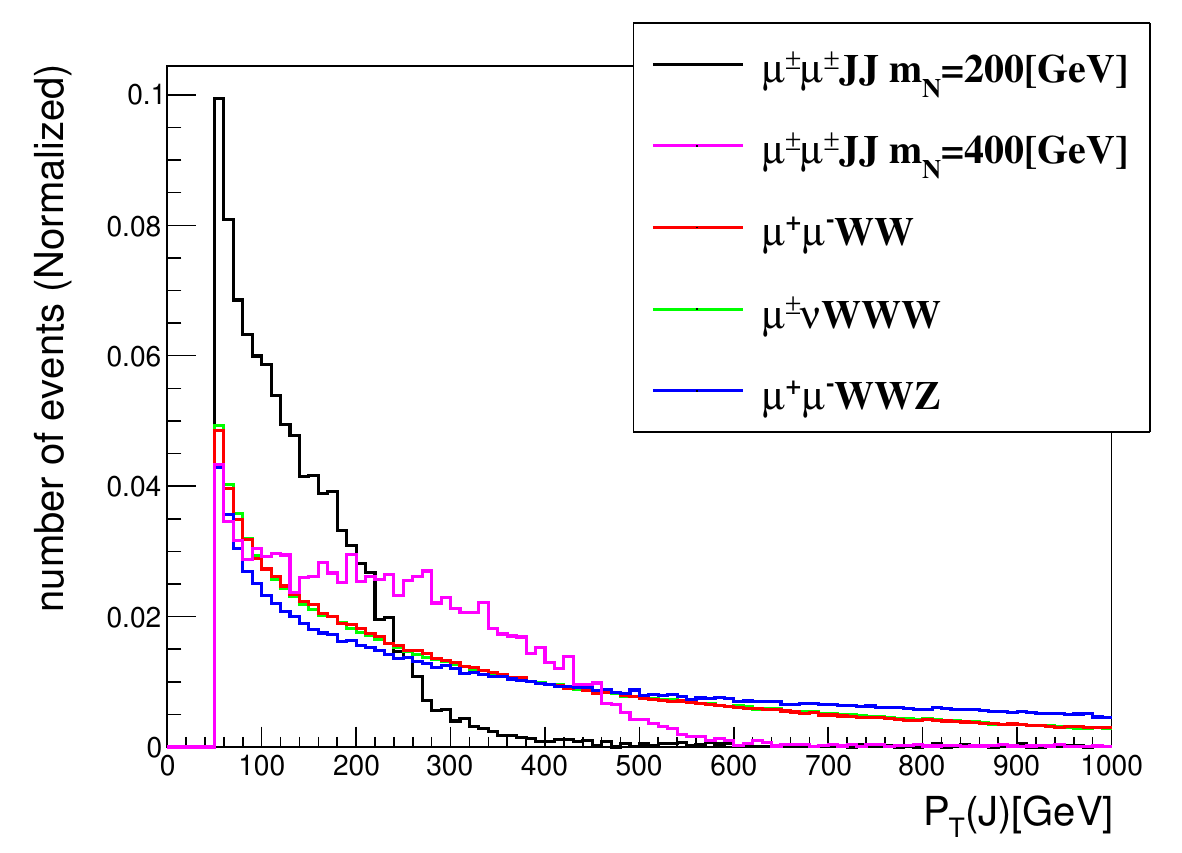}
		\includegraphics[width=0.45\linewidth]{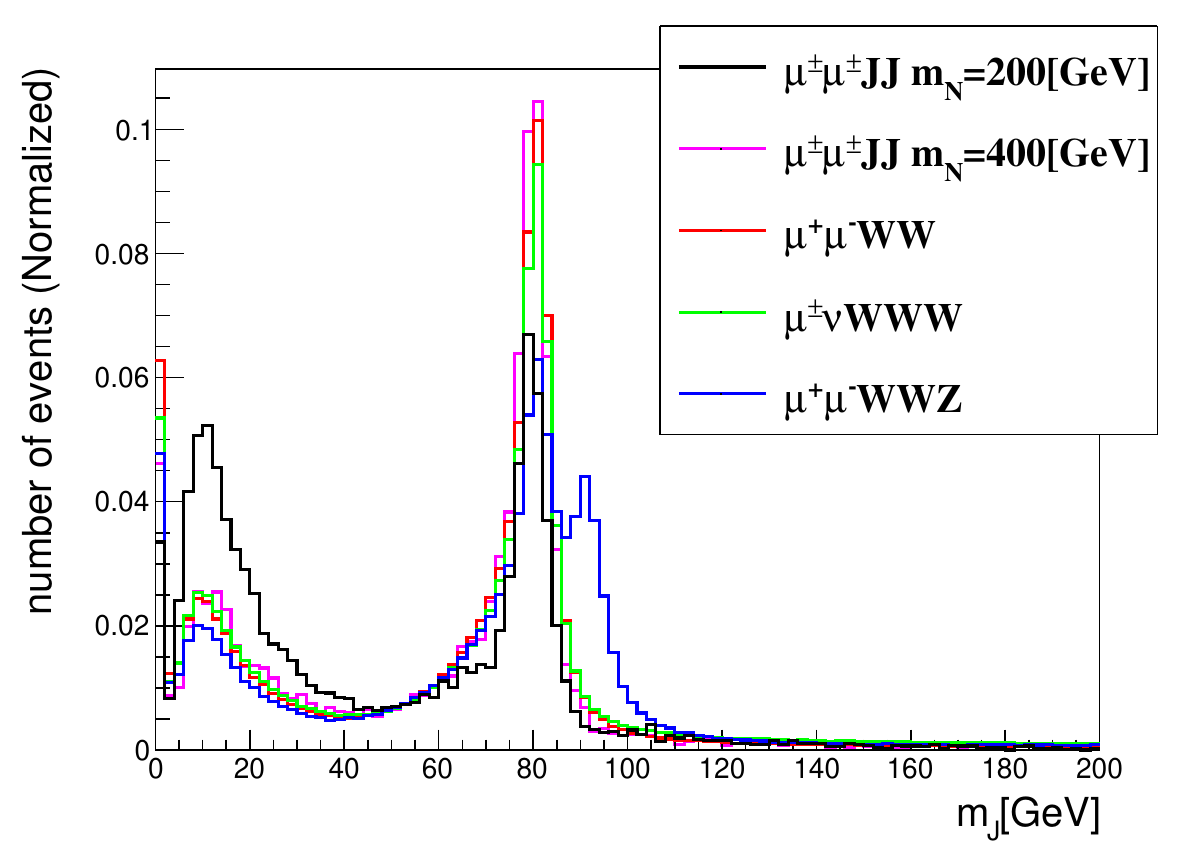}
		\includegraphics[width=0.45\linewidth]{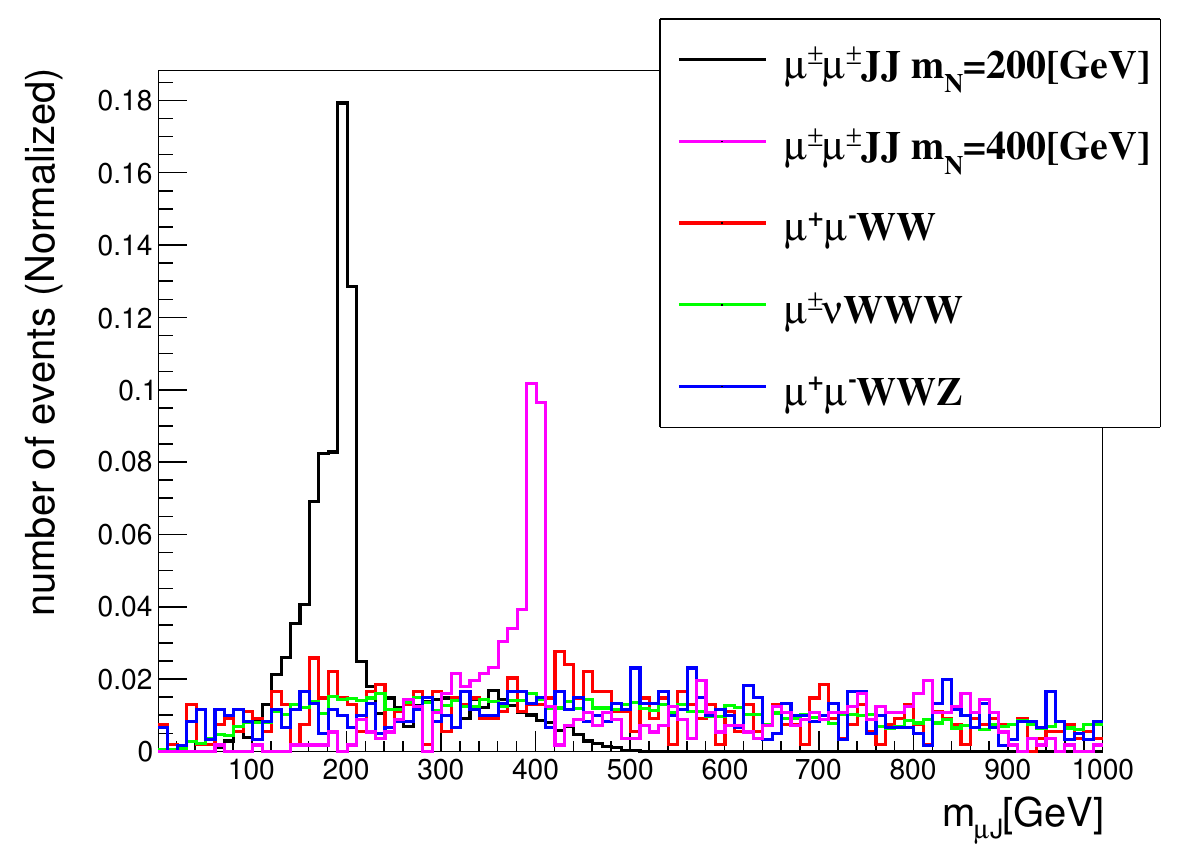}
	\end{center}
	\caption{Normalized distributions of muon transverse momentum $P_T(\mu)$ (up-left panel), fat-jet transverse momentum $P_T(J)$ (up-right panel), invariant mass of fat-jet $m_{J}$ (down-left panel), and invariant mass of muon and fat-jet $m_{\mu{J}}$ (down-right panel)  for the signal and corresponding backgrounds at the 3 TeV muon collider. We choose two benchmark points with $m_H/3=m_N=200~\rm{GeV}$ and $m_H/3=m_N=400~\rm{GeV}$ with $\sin\alpha=0.1$.}
	\label{fig8}
\end{figure}

After the simulation of signal and backgrounds events, the same pre-selection cuts as in Section \ref{Sec:SM} are firstly applied 
\begin{align}
	|\eta(\mu)|<2.5, P_T(J)>50~{\rm{GeV}}, |\eta(J)|<2.5. 
\end{align}
Distributions of some variables for the signal and backgrounds at the 3 TeV muon collider are shown in Figure \ref{fig8}. The benchmark points are $m_H/3=m_N=200~\rm{GeV}$ and $m_H/3=m_N=400~\rm{GeV}$ with $\sin\alpha=0.1$.

As shown in Figure \ref{fig7}, the typical cross section of $H\to \mu^\pm \mu^\pm JJ$ is less than 1 fb, which is smaller than the cross section of $h\to \mu^\pm \mu^\pm JJ$ with the same value of $\sin\alpha$. In the upper panels of Figure \ref{fig8}, we find that the transverse momentum of the muon $P_T(\mu)$ tends to be larger when $m_N$ increases. For the SM backgrounds,  there are peaks around $P_T(\mu)\sim850$ GeV. To keep more signal events, we have changed the selection cuts on muons as
\begin{align}
	N_\mu^{\pm}=2, 20~{\rm{GeV}}<P_T(\mu)<m_N.
\end{align}
Here, the lower cut on $P_T(\mu)$ is selected heuristically. The best cut on the lower value of $P_T(\mu)$ indeed depends on $m_N$. We do not vary it in the analysis, because the difference of the significance is relatively small between the best and fixed cut on lower value of $P_T(\mu)$.

In Table \ref{Tab02}, we summarize the cut flow for the $H\to{\mu^\pm}{\mu^\pm}JJ$ signature and backgrounds. The benchmark point of the signal is fixed as $m_H/3=m_N=200$ GeV and $\sin\alpha=0.1$. The acceptance efficiency of the same-sign dimuon cuts is about $0.5\sim0.6$, which is much larger than it of the $h\to \mu^\pm\mu^\pm JJ$ signature. Meanwhile, the larger range of $P_T(\mu)$ leads to larger SM backgrounds than the analysis in Section \ref{Sec:SM}. After the selection cuts on the same-sign dimuon, the total cross section of backgrounds is at the order of $\mathcal{O}(10^{-1})$ fb at the 3 TeV muon collider, which is in the same order as the signal. Therefore, more advanced cuts are expected.

On the other hand, the distributions of the transverse momentum of the fat-jet $P_T(J)$ from $H\to \mu^\pm \mu^\pm JJ$ are different from $h\to \mu^\pm \mu^\pm JJ$. For a larger heavy neutral lepton mass, the transverse momentum of the fat-jet $P_T(J)$ also tends to be larger, so we modified the cuts on fat-jets as
\begin{align}
	N_J\geq1,50{\rm{GeV}}<P_T(J)<m_N.
\end{align}
 Considering the fact that the fat-jets in this section are from the on-shell $W$ decay. We then require the invariant mass of fat-jets as
\begin{equation}
	50~\text{GeV}<m_J<120 ~\text{GeV},
\end{equation}
which is based on the distribution of $m_J$ in Figure \ref{fig8}. Finally, the reconstructed mass of heavy neutral lepton is also required in the range of $0.8m_N<m_{\mu{J}}<1.2m_N$.

\begin{table}
	\begin{center}
		\begin{tabular}{c | c | c | c | c} 
			\hline
			\hline
			$\sigma(fb)$&${\mu^\pm}{\mu^\pm}JJ$&${\mu^+}{\mu^-}WW$&$\mu^\pm{\nu}WWW$&$\mu^+\mu^-WWZ$ \\
			\hline
			Pre-selection &0.24(0.67) &28(13) &5.8(11) &0.84(1.3) \\
			\hline
			muon cuts  &0.14(0.32) & $2.0\times10^{-2}$($7.6\times10^{-4}$) & $9.7\times10^{-2}$($1.3\times10^{-2}$)&  $9.9\times10^{-4}$($1.8\times10^{-4}$) \\
			\hline
			fat-jet cuts &0.11(0.24)  & $7.5\times10^{-3}$($1.0\times10^{-4}$) & $5.9\times10^{-2}$($3.5\times10^{-3}$) & $5.2\times10^{-4}$($4.3\times10^{-5}$)\\
			\hline
			$m_{\mu{J}}$ cut  &0.10(0.21)  & $2.4\times10^{-3}$($2.6\times10^{-5}$) & $2.3\times10^{-2}$($1.1\times10^{-3}$) & $1.9\times10^{-4}$($7.6\times10^{-6}$)\\
			\hline
			\hline
			Significance & 14(134)  & \multicolumn{2}{|c|}{Total Background} & $2.6\times10^{-2}$($1.1\times10^{-3}$) \\
			\hline
		\end{tabular}
	\end{center}
	\caption{Cut flow table for the signal and three dominant backgrounds process with $m_H/3=m_N=200~\rm{GeV}$ and $\sin\alpha=0.1$ at the 3TeV (10TeV) muon collider. The significance is calculated with an integrated luminosity of 1 (10) ab$^{-1}$ for the 3TeV (10TeV) muon collider.
	\label{Tab02}}
\end{table}

The cross section of $H\to \mu^\pm \mu^\pm JJ$ of the benchmark point passing all selection cuts is 0.10(0.21) fb at the 3(10)TeV muon collider with $\sin\alpha=0.1$, while the total cross section of backgrounds is $2.6\times10^{-2}(1.1\times10^{-3})$ fb. The total acceptance efficiency of $H\to \mu^\pm \mu^\pm JJ$ signature is about 0.42(0.31) at the 3(10) TeV muon collider,  which is much larger than the acceptance efficiency of $h\to \mu^\pm \mu^\pm JJ$. So, although the  initial production cross section of $H\to \mu^\pm \mu^\pm JJ$ is much smaller than it of $h\to \mu^\pm \mu^\pm JJ$, the final cross sections after all cuts are at the same order for these two signature. However, with larger upper limits on the transverse momentum of muon and fat-jet, the final cross section of the total SM backgrounds is about two orders of magnitudes larger than the results in Section \ref{Sec:SM}. With an integrated luminosity of $1~\rm{ab^{-1}}(10~\rm{ab^{-1}})$, we can get about 100(2100) signal events and 26(11) background events at the 3(10) TeV muon collider. Then the significance $S=14(134)$ is achieved. To reach the $5\sigma$ discovery limit, the required luminosity is about 127(14) fb$^{-1}$. Therefore, the lepton number violation heavy Higgs decay signature is also promising at the TeV scale muon collider.

\begin{figure}
	\begin{center}
		\includegraphics[width=0.45\linewidth]{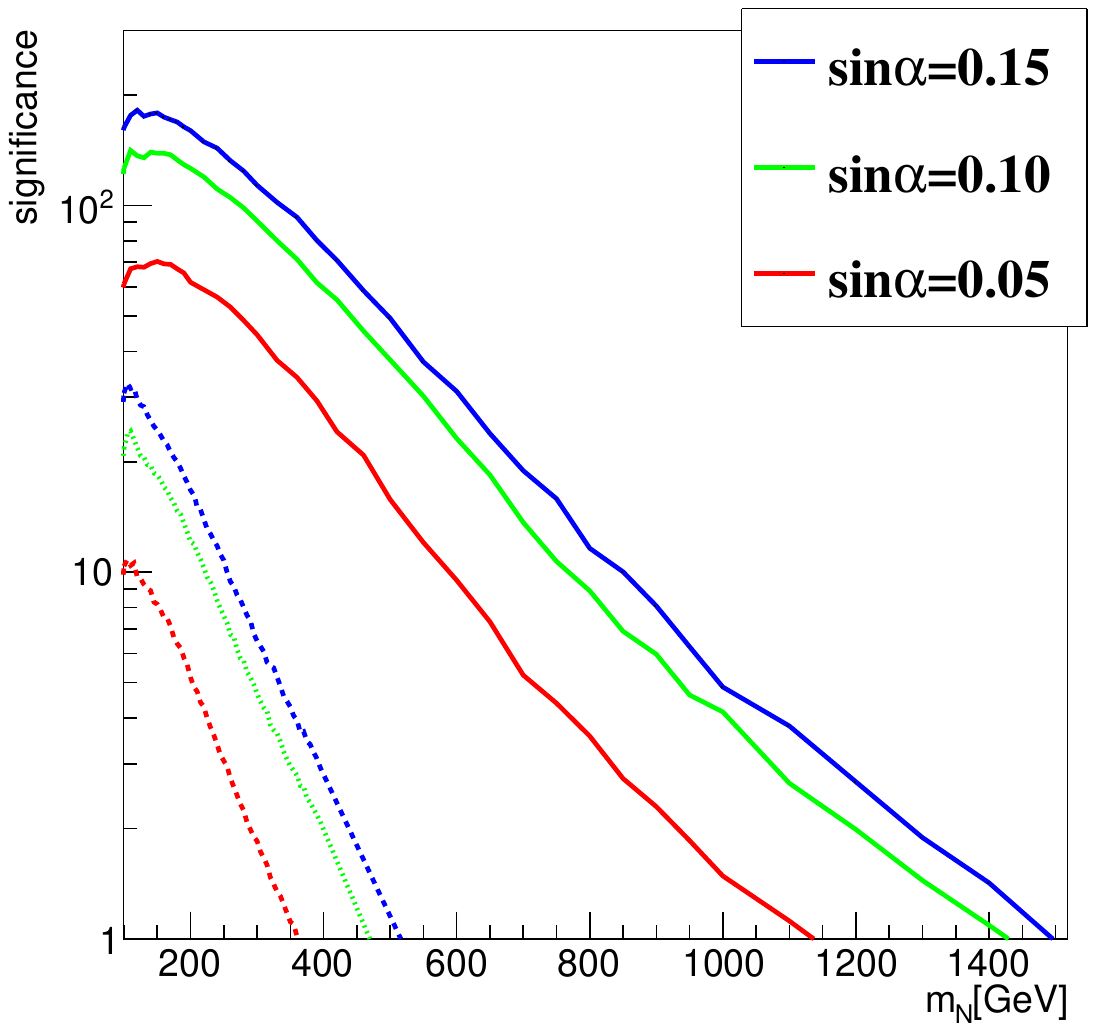}
		\includegraphics[width=0.45\linewidth]{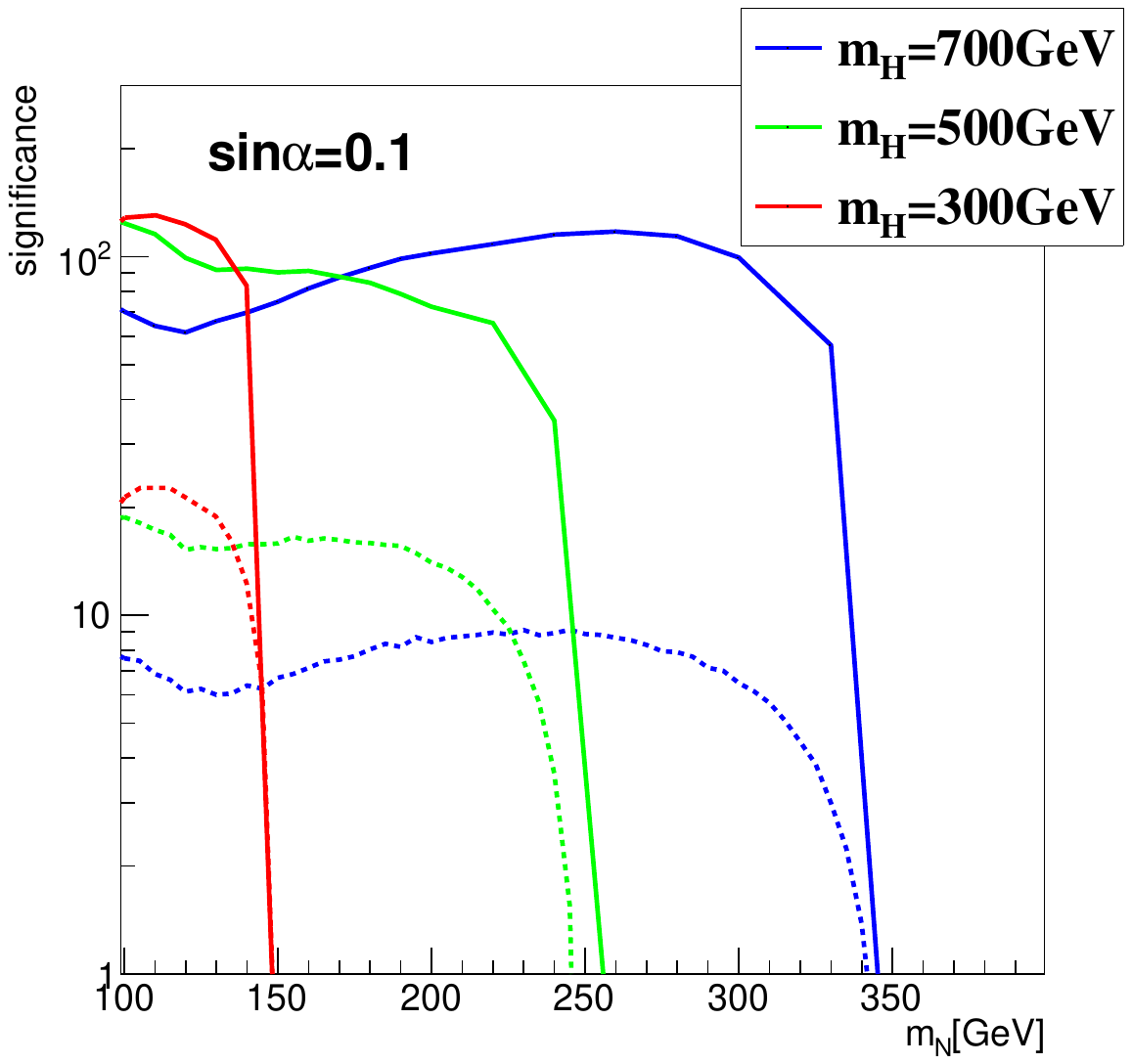}
		\includegraphics[width=0.45\linewidth]{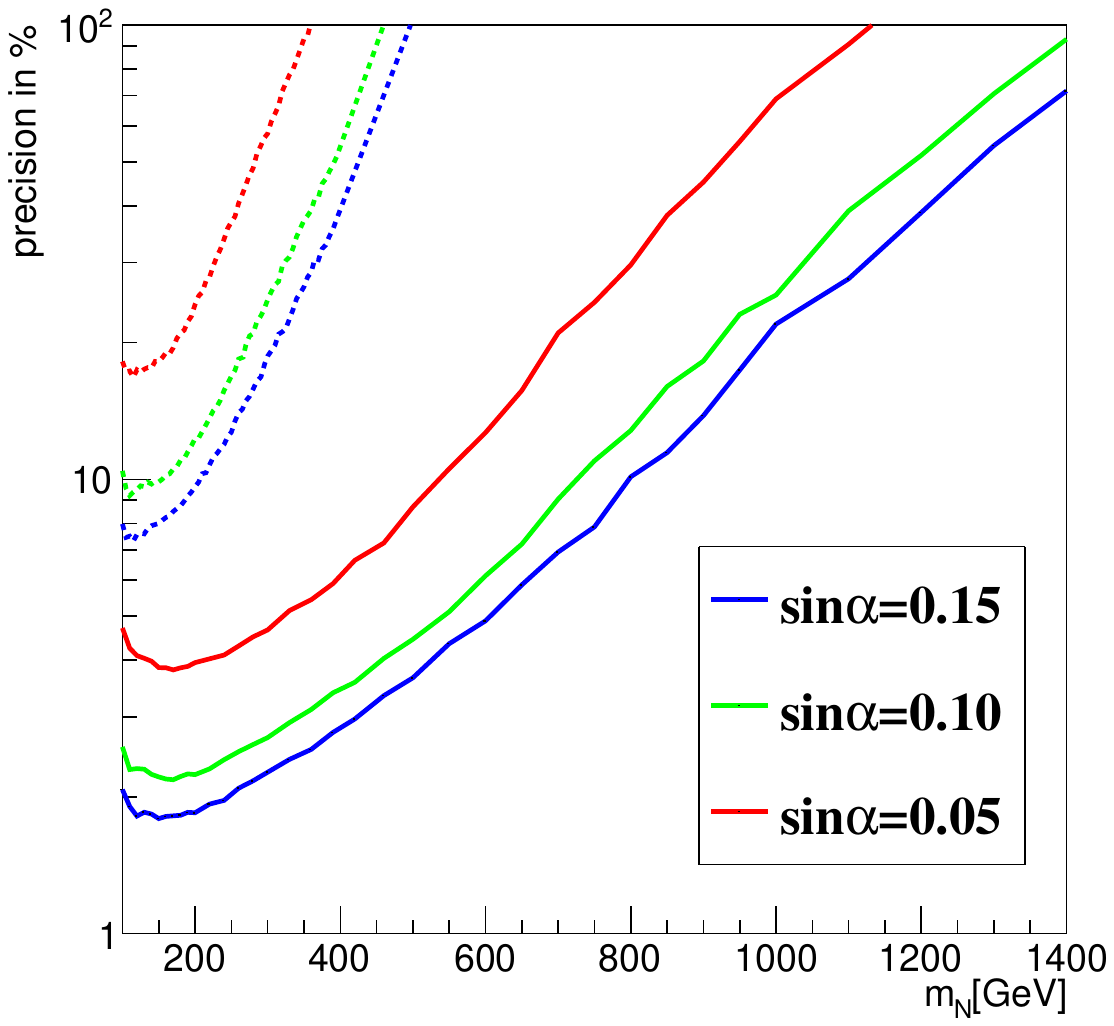}
		\includegraphics[width=0.45\linewidth]{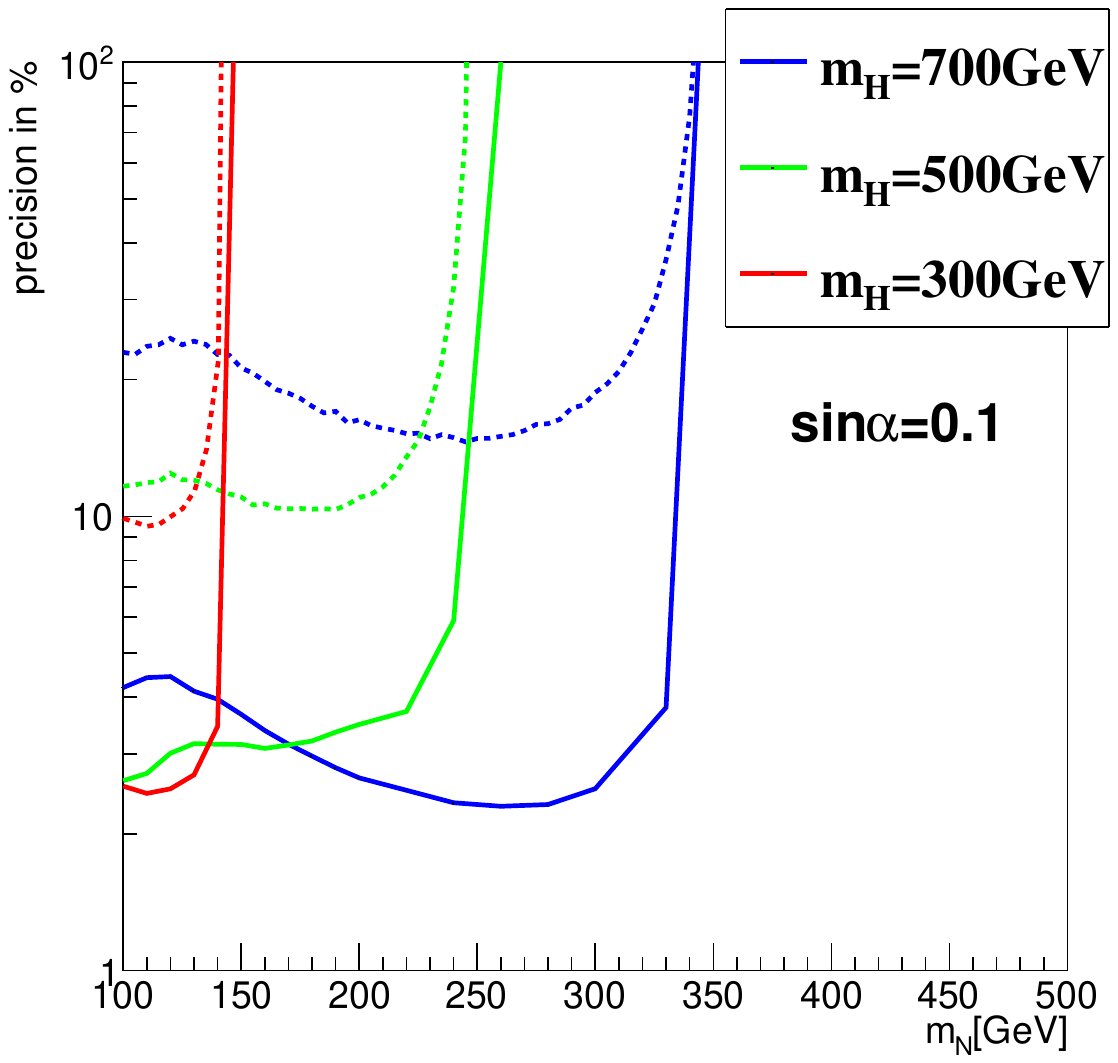}
	\end{center}
	\caption{The significance (upper panels) and precision (lower panels) of the $H\to\mu^{\pm}\mu^{\pm}JJ$ signature. In the left panels we have fixed the mass relation $m_N=m_H/3$. The blue, green, and red lines are the results for $\sin{\alpha}=0.15,0.1$, and $0.05$, respectively. 
	In the right panels we have fixed $\sin{\alpha}=0.1$. The blue, green, and red lines are the results for $m_{H}=$300 GeV, 500 GeV, 700GeV.  The dashed (solid) lines are for 3 (10) TeV muon collider with an integrated luminosity of $L=1(10)~\text{ab}^{-1}$. }
	\label{fig9}
\end{figure}

Based on the selection cuts in Table \ref{Tab02}, we then explore the significance of the $H\to \mu^\pm \mu^\pm JJ$ signature at the muon collider. The results are shown in Figure \ref{fig9}. In the left-upper panel, we show the significance of the $H\to \mu^\pm \mu^\pm JJ$ signature as a function of $m_N$ for $\sin\alpha=0.15,0.10$, and 0.05 respectively, where we have fixed the mass relation $m_N=m_H/3$. The significance decreases as $m_N$ increases. On the 3TeV muon collider with $1~\text{ab}^{-1}$ integrated luminosity, we can reach 5 $\sigma$ discovery for $m_{N}\lesssim 330$ GeV and $\sin\alpha=0.15$. Lowering $\sin\alpha$ to 0.05, the significance still exceeds 5 $\sigma$ when $m_N\lesssim 200$ GeV.  Increasing the collision energy to 10 TeV with 10 ab$^{-1}$ data could extend the discovery limit up to $m_N\lesssim1$ TeV with $\sin\alpha=0.1$. The precision at 3 TeV is about 10\% $\sim$ 20\% when $m_N\lesssim300$ GeV, and it is less than 10\% at 10 TeV when $m_N\lesssim800$ GeV and $\sin\alpha\gtrsim0.1$.

In the right-upper panel of Figure \ref{fig9}, we show the significance of the $H\to \mu^\pm \mu^\pm JJ$ signature as a function of $m_N$ for $m_H=$300 GeV, 500 GeV  and 700 GeV respectively, where we have fixed the Higgs mixing $\sin\alpha=0.1$. For $m_H=300$ GeV, the sensitivity reaches the maximum value of 23 $\sigma$ around $m_N=110$ GeV at the 3 TeV muon collider, and such result can reach to 130 $\sigma$ around $m_N=110$~GeV at the 10 TeV muon collider,. Meanwhile, for $m_H=500$ GeV and 700 GeV, we find that there exists a local minimum value of significance around $m_N\sim135$ GeV. And the 5 $\sigma$ discovery could be realized when $m_N\lesssim 230$~GeV with $m_H=500$ GeV or  $m_N\lesssim 320$ GeV with $m_H=700$ GeV at the 3 TeV muon collider, and $m_N\lesssim 240$~GeV with $m_H=500$ GeV or  $m_N\lesssim 340$ GeV with $m_H=700$ GeV at the  10 TeV muon collider. Therefore, the 10 TeV muon collider is expected to probe most regions below the kinematic threshold $m_N\lesssim m_H/2$. We find a precision of $\mathcal{O}(10)\%$ at the 3 TeV muon collider and $\mathcal{O}(3)\%$ at the 10 TeV muon collider, respectively.

\begin{figure}
	\begin{center}
		\includegraphics[width=0.45\linewidth]{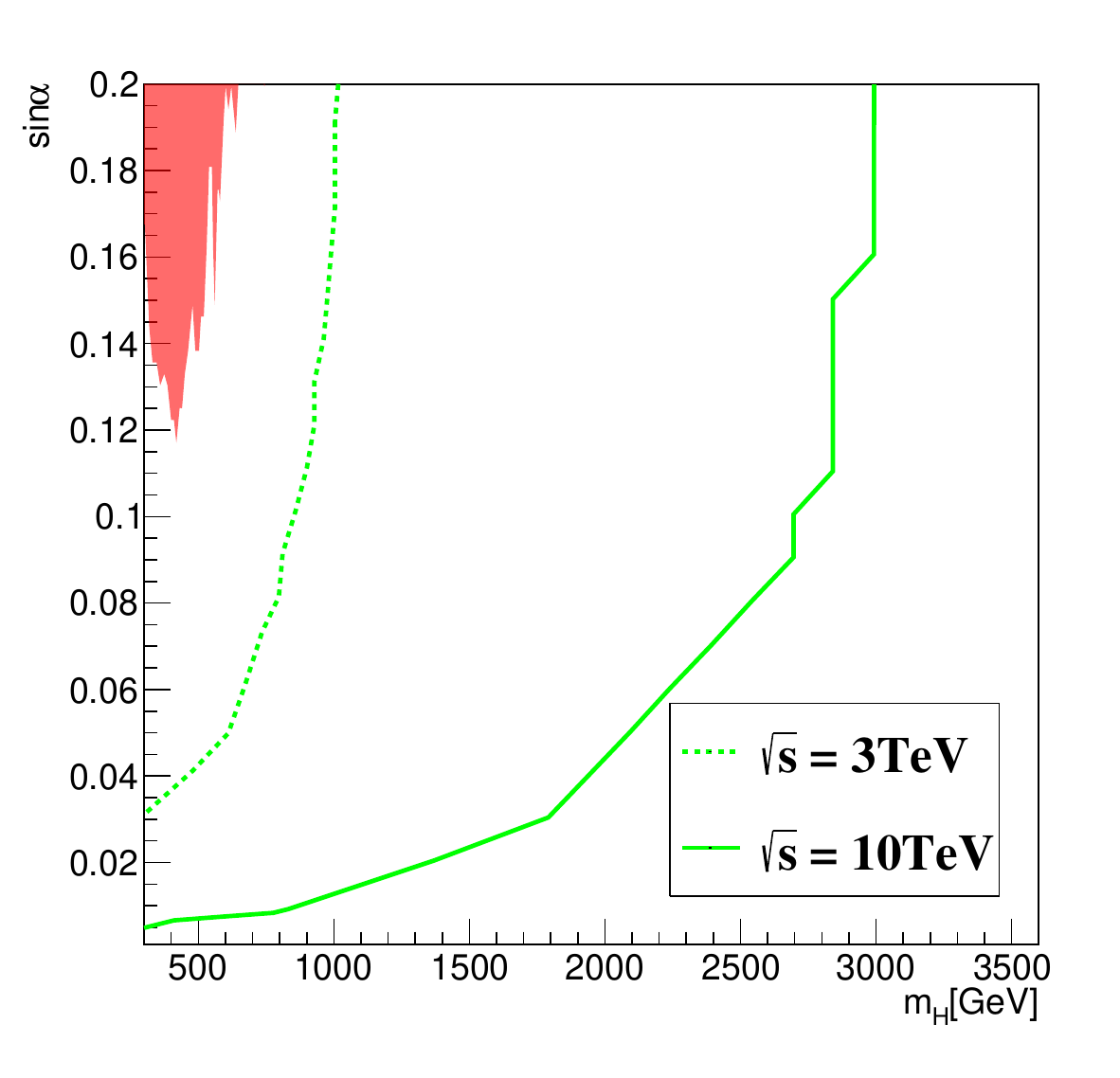}
		\includegraphics[width=0.45\linewidth]{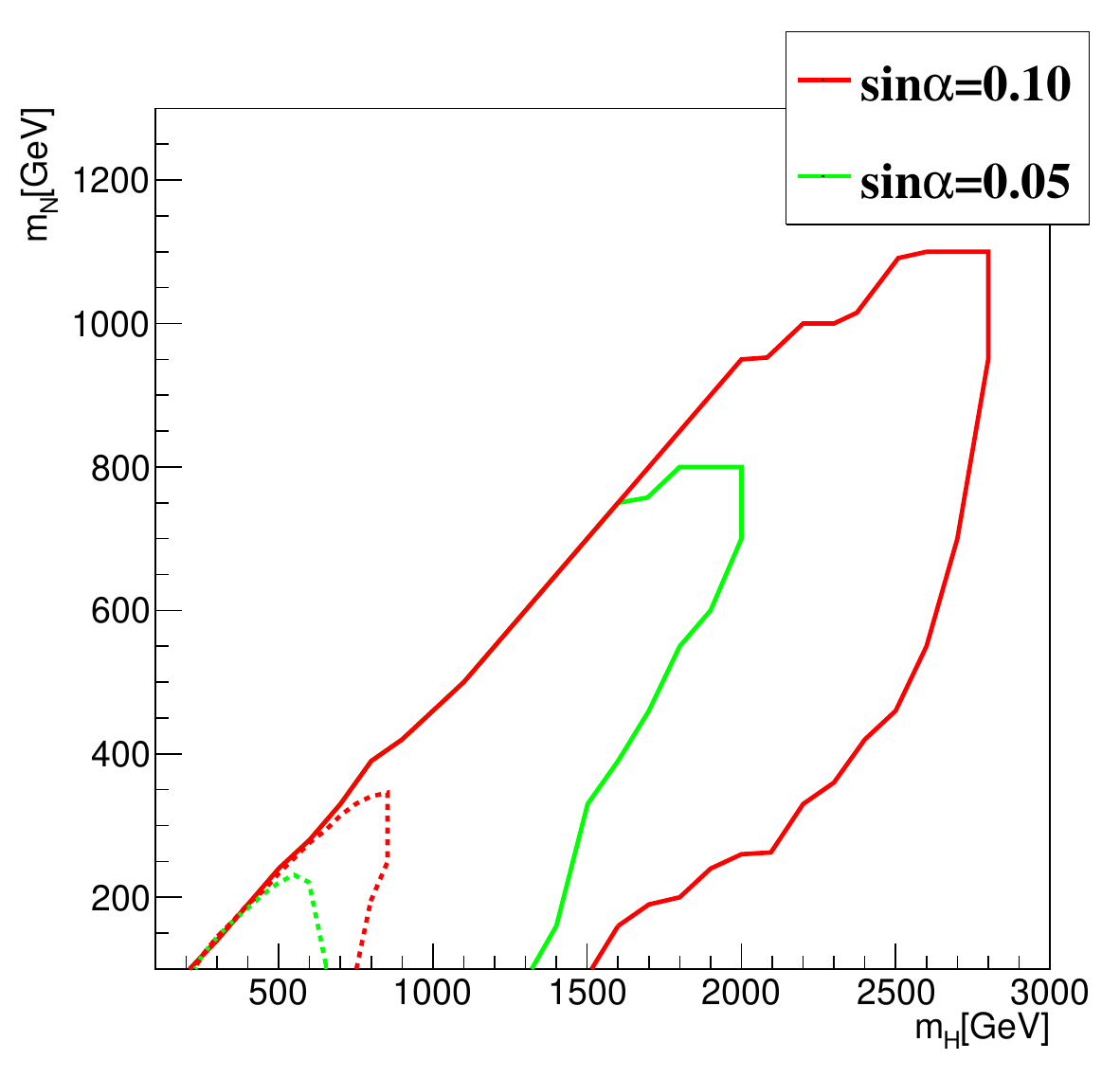}
	\end{center}
	\caption{The discovery reach of the $H\to\mu^{\pm}\mu^{\pm}JJ$ signature.  In the left panel we have set $m_N=m_H/3$. The red region is excluded by direct searches of heavy Higgs \cite{Lane:2024vur}. In the right panel, the red and green lines are the results for $\sin\alpha=0.1$ and $0.05$. }
	\label{fig10}
\end{figure}

The $5\sigma$ discovery region in the $m_N-\sin\alpha$ plane with $m_N=m_H/3$ is shown in the  left panel of Figure \ref{fig10}. At the 3 TeV muon collider with $1~\text{ab}^{-1}$ data, the $H\to\mu^{\pm}\mu^{\pm}JJ$ signature can be detected with the Higgs mixing parameter as small as $\sin\alpha\simeq 0.03$. According to the results in Figure \ref{fig7}, the cross section of $H\to\mu^\pm\mu^\pm JJ$ is heavily suppressed when $\sin\alpha\lesssim0.01$, thus it becomes less promising. Due to the larger cross section and integrated luminosity, the 10 TeV muon collider could discover the Higgs mixing parameter as small as $\sin\alpha\simeq 0.005$ for heavy Higgs around the electroweak scale. With larger singlet scalar mass, the value of Higgs mixing parameter $\sin\alpha$ increases for 5 $\sigma$ discovery. Typically, for $\sin\alpha=0.15$, we might probe $m_N\lesssim 330$~GeV and $m_N\lesssim 950$~GeV for 3 TeV muon collider and 10 TeV muon collider separately.

In the right panel of Figure \ref{fig10}, we show the sensitivity region in the $m_{H}-m_N$ plane with fixed  $\sin\alpha$ of 0.1 and 0.05. At the 3 TeV muon collider, a large parameter space can be detected for $230~{\rm{GeV}}<m_H<880~\rm{GeV}$ and $m_N<340~\rm{GeV}$ with $\sin\alpha=0.1$. Of course, the sensitive region heavily depends on the Higgs mixing parameter. For $\sin\alpha>0.1$, although certain region with $m_H\lesssim600$ GeV is already excluded by current direct searches of heavy Higgs, $m_H$ around TeV scale is also promising with such large $\sin\alpha$. When $\sin\alpha=0.05$, the promising region becomes $230~{\rm{GeV}}<m_H<640~\rm{GeV}$ and $m_N<220~\rm{GeV}$, which is clearly smaller than the scenario with $\sin\alpha=0.1$. Increasing the collision energy to 10 TeV will greatly enlarge the discovery region. For instance, the parameter space can be detected for $220~{\rm{GeV}}<m_H<2800~\rm{GeV}$ and $m_N<1100~\rm{GeV}$ with $\sin\alpha=0.1$, as well as $220~{\rm{GeV}}<m_H<2000~\rm{GeV}$ and $m_N<800~\rm{GeV}$ for $\sin\alpha=0.05$.

%%%%%%%%%%%%%%%%%%%%%%%%%%%%%%%%
\section{Conclusion}\label{SEC:CL}

Tiny neutrino mass could be explained by the type-I seesaw mechanism by introducing heavy neutral lepton $N$. In the scalar singlet $S$ extension of the type-I seesaw, the Majorana mass term of heavy neutral lepton originates from the Yukawa interaction $\frac{y_N}{2} S\bar{N}^c N$, which generates the Majorana mass as $m_N=y_N v_s/\sqrt{2}$ after $S$ develops vacuum expectation value. Due to mixing between scalars, both the SM Higgs $h$ and the additional heavy Higgs $H$ couple to the heavy neutral lepton. The lepton number violation decays as $h/H\to NN\to \mu^\pm \mu^\pm 4j$ provides an appealing pathway to probe the origin of heavy neutral lepton mass.

In this paper, we study the lepton number violation Higgs decay signature at the TeV-scale muon collider. Dominant by the vector boson fusion channel, the explicit signal process at muon collider is $\mu^+\mu^-\to \nu_\mu \bar{\nu}_\mu  h/H \to \nu_\mu \bar{\nu}_\mu NN $ followed by $ N \to \mu^\pm jj$, where the two jets from $W$ boson decay are treated as one fat-jet $J$. For illustration, we further assume that the heavy neutral lepton $N$ preferentially couples to muon via mixing with light neutrinos. For the  SM Higgs decay signature $h\to \mu^\pm\mu^\pm JJ$, the mass of heavy neutral lepton should satisfy $m_N<m_h/2$, so the fat-jets $J$ are from off-shell $W$ decay. On the other hand, we assume that the mass of heavy neutral lepton is larger than $W$ boson mass when considering the  heavy Higgs decay signature $H\to \mu^\pm \mu^\pm JJ$, so the fat-jets $J$ are from on-shell $W$ decay in this scenario. In this way, these two kinds of lepton number violation signatures can be distinguished.

The lepton number violation SM Higgs decay signature depends on the mass of heavy neutral lepton $m_N$ and Higgs mixing parameter $\sin\alpha$, while the lepton number violation heavy Higgs decay signature further depends on the mass of heavy Higgs mass $m_H$. We then explore the sensitivity of these signatures at the 3 TeV and 10 TeV muon collider. With a relatively clean SM background, we perform a cut-based analysis to reduce the backgrounds. For the  SM Higgs decay signature $h\to {\mu^\pm}{\mu^\pm}JJ$, the sensitivity could reach the 5 $\sigma$ discovery limits when $m_N\in[10,58]~\rm{GeV}$ with $\sin\alpha=0.2$ at the 3 TeV muon collider with an integrated luminosity of $L=1~\text{ab}^{-1}$. The limit of the mixing parameter $\sin\alpha$ can be as small as $\sin\alpha=0.05$ when $m_N=45$~GeV, which corresponds to the branching ratio of $h\to NN$ as $4.1\times10^{-3}$. At the 10 TeV muon collider with an integrated luminosity of $L=10~\text{ab}^{-1}$, the discovery region is enlarged due to a larger signal cross section. For instance, the discovery limit of the mixing parameter $\sin\alpha$ can be down to $\sin\alpha=0.009$ when $m_N=45$~GeV, which corresponds to the branching ratio of $h\to NN$ as $1.3\times10^{-4}$.

For the heavy Higgs decay signature $H\to {\mu^\pm}{\mu^\pm}JJ$, a larger heavy Higgs mass usually requires a larger Higgs mixing parameter to reach the 5 $\sigma$ discovery limit. For example, at the 3 TeV muon collider with an integrated luminosity of $L=1~\text{ab}^{-1}$, we may discover $m_H\lesssim 1$~TeV with $m_N=m_H/3$ for $\sin\alpha=0.15$. Meanwhile the $H\to\mu^{\pm}\mu^{\pm}JJ$ signature can be detected with the Higgs mixing parameter as small as $\sin\alpha\simeq 0.03$ when $m_N=100$ GeV. With $1~\text{ab}^{-1}$ data, a large parameter space of the signature $H\to {\mu^\pm}{\mu^\pm}JJ$ can be detected for $230~{\rm{GeV}}<m_H<880~\rm{GeV}$ and $m_N<340~\rm{GeV}$ with $\sin\alpha=0.1$. At the 10 TeV muon collider with an integrated luminosity of $L=10~\text{ab}^{-1}$, we may discover $m_H\lesssim 2.85$~TeV with $m_N=m_H/3$ for $\sin\alpha=0.15$, meanwhile the $H\to\mu^{\pm}\mu^{\pm}JJ$ signature can be detected with the Higgs mixing parameter down to $\sin\alpha\simeq 0.005$ when $m_N=100$ GeV. With $10~\text{ab}^{-1}$ data, a large parameter space of the signature $H\to {\mu^\pm}{\mu^\pm}JJ$ can be detected for $220~{\rm{GeV}}<m_H<2800~\rm{GeV}$ and $m_N<1100~\rm{GeV}$ with $\sin\alpha=0.1$.

\section{Acknowledgments}

This work is supported by the National Natural Science Foundation of China under Grant No. 12375074 and 12175115, Natural Science Foundation of Shandong Province under Grant No. ZR2022MA056 and ZR2024QA138, the Open Project of Guangxi Key Laboratory of Nuclear Physics and Nuclear Technology under Grant No. NLK2021-07, University of Jinan Disciplinary Cross-Convergence Construction Project 2024 (XKJC-202404).

%%%%%%%%%%%%%%%%%%%%%%%%%%%%%

\end{document}